\documentclass[fleqn,usenatbib]{mnras}

\usepackage{newtxtext,newtxmath}

\usepackage[T1]{fontenc}

\DeclareRobustCommand{\VAN}[3]{#2}
\let\VANthebibliography\thebibliography
\def\thebibliography{\DeclareRobustCommand{\VAN}[3]{##3}\VANthebibliography}

\usepackage{graphicx}	
\usepackage{amsmath}	
\usepackage{hyperref}
\usepackage{color}
\definecolor{patriarch}{rgb}{0.5, 0.0, 0.5}

\usepackage{natbib}

\title[Dust Attenuation and Morphology]{Big, Dusty Galaxies in Blue Jay: Insights into the Relationship Between Morphology and Dust Attenuation at Cosmic Noon}

\author[G. Maheson et al.]{Gabriel Maheson,$^{1,2}$\thanks{E-mail: gm695@cam.ac.uk}
Sandro Tacchella,$^{1,2}$
Sirio Belli,$^{3,4}$
Minjung Park,$^{5}$ 
A. Lola Danhaive,$^{1,2}$ 
\newauthor Letizia Bugiani,$^{3,4}$
Rebecca Davies,$^{6,7}$
Razieh Emami,$^{5}$
Amir H. Khoram,$^{3,4}$
Laurence Lam,$^{2}$ 
Joel Leja,$^{8,9,10}$
\newauthor Trevor Mendel$^{7,11}$ and 
Erica June Nelson$^{12}$
\\
$^{1}$Kavli Institute for Cosmology, University of Cambridge, Madingley Road, Cambridge CB3 0HA, UK\\
$^{2}$Cavendish Laboratory, University of Cambridge, 19 J. J. Thomson Ave., Cambridge CB3 0HE, UK \\
$^{3}$Dipartimento di Fisica e Astronomia, Università di Bologna, Via Gobetti 93/2, I-40129, Bologna, Italy\\
$^{4}$INAF, Astrophysics and Space Science Observatory Bologna, Via P. Gobetti 93/3, I-40129 Bologna, Italy\\
$^{5}$ Center for Astrophysics $\vert$ Harvard \& Smithsonian, 60 Garden Street, Cambridge, MA 02138, USA\\
$^{6}$Centre for Astrophysics and Supercomputing, Swinburne University of Technology, Hawthorn, Victoria, Australia\\
$^{7}$5ARC Centre of Excellence for All Sky Astrophysics in 3 Dimensions (ASTRO 3D), Australia\\
$^{8}$ Department of Astronomy \& Astrophysics, The Pennsylvania State University, University Park, PA 16802, USA\\
$^{9}$Institute for Computational \& Data Sciences, The Pennsylvania State University, University Park, PA 16802, USA\\
$^{10}$Institute for Gravitation and the Cosmos, The Pennsylvania State University, University Park, PA 16802, USA\\
$^{11}$Research School of Astronomy and Astrophysics, Australian National University, Canberra, ACT, Australia\\
$^{12}$Department for Astrophysical and Planetary Science, University of Colorado, Boulder, CO 80309, USA\\
}

\date{Accepted XXX. Received YYY; in original form ZZZ}

\pubyear{2025}

\begin{document}
\label{firstpage}
\pagerange{\pageref{firstpage}--\pageref{lastpage}}
\maketitle

\begin{abstract}
The dust attenuation of galaxies is highly diverse and closely linked to stellar population properties and the star–dust geometry, yet its relationship to galaxy morphology remains poorly understood. We present a study of 141 galaxies ($9<\log(\rm M_{\star}/\rm M_{\odot})<11.5$) at $1.7<z<3.5$ from the Blue Jay survey combining deep JWST/NIRCam imaging and $R\sim1000$ JWST/NIRSpec spectra. Using \texttt{Prospector} to perform a joint analysis of these data with non-parametric star-formation histories and a two-component dust model with flexible attenuation laws, we constrain stellar and nebular properties. We find that the shape and strength of the attenuation law vary systematically with optical dust attenuation ($A_V$), stellar mass, and star formation rate (SFR). $A_V$ correlates strongly with stellar mass for starbursts, star-forming galaxies and quiescent galaxies. The inclusion of morphological information tightens these correlations: attenuation correlates more strongly with stellar mass and SFR surface densities than with the global quantities. The Balmer decrement-derived nebular attenuation for 67 of these galaxies shows consistent trends with the stellar continuum attenuation. We detect a wavelength-dependent size gradient: massive galaxies ($\rm M_{\star}\gtrsim 10^{10}~M_{\odot}$) appear $\sim30\%$ larger in the rest-optical than in the rest-NIR, driven by central dust attenuation that flattens optical light profiles. Lower-mass systems exhibit more diverse size ratios, consistent with either inside-out growth or central starbursts. These results demonstrate that dust attenuation significantly alters observed galaxy structure and highlight the necessity of flexible attenuation models for accurate physical and morphological inference at cosmic noon.
\end{abstract}




\section{Introduction}

Evolved asymptotic giant branch (AGB) stars and supernovae are the primary sources of dust in galaxies \citep{draine_physics_2011, gall_production_2011, sarangi_dust_2018, micelotta_dust_2018, schneider_formation_2024}, and so the dust content of a galaxy is closely linked to the stellar population. This dust, with grains ranging in size from $5-250$ nm \citep{weingartner_dust_2001}, has been observed to have features implying a carbon-based composition \citep[e.g.,][]{stecher_graphite_1965, draine_physics_2011, witstok_carbonaceous_2023, schneider_formation_2024, lin_polycyclic_2025} , as well as features implying a silicate-based composition \citep[e.g.,][]{gillett_anisotropy_1975, draine_interstellar_2003, kemper_absence_2004, salim_dust_2020}. Although the exact chemical composition of the dust is not fully constrained, it significantly impacts a galaxy's spectral energy distribution (SED) through the absorption, scattering and re-emission of the light from the stars, ionised nebulae and any other source of light. The absorption and scattering of light by dust is known as dust attenuation and depends on the composition of the dust and the relative geometry of the dust and sources of light. Dust attenuation is wavelength-dependent, with dust being more effective at scattering and absorbing the light at shorter wavelengths (e.g., rest-frame UV) than at longer wavelengths (e.g., rest-frame optical to rest-frame near-infrared). At longer wavelengths, the dust starts to emit radiation approximating a black-body-spectrum of one or more characteristic temperatures between the rest-frame near-infrared (NIR) and the rest-frame far-infrared, depending on the dust temperature \citep{draine_interstellar_2003, da_cunha_new_2010, orellana_molecular_2017}. 

The wavelength dependence of the dust attenuation, or the dust attenuation law, has been studied for decades for nearby galaxies \citep[e.g.,][]{calzetti_dust_2000, wild_empirical_2011, salim_dust_2018}, to some of the first galaxies which formed in the first billion years after the Big Bang \citep[e.g.,][]{gallerani_extinction_2010, stratta_is_2011, laporte_dust_2017, popping_dust_2017, mascia_dust_2021, wang_luminous_2021, inami_alma_2022, witstok_carbonaceous_2023, markov_dust_2023, markov_evolution_2024}. The dust within our own Galaxy has been measured extensively; however, its effect on stellar light is generally modelled as a screen of dust in front of the star, whereby the dust acts to simply scatter the light out of the line of sight. This effect is called dust extinction, which has been constrained well for the Milky Way \citep[MW;][]{cardelli_relationship_1989}, the Large Magellanic Cloud (LMC) and the Small Magellanic Cloud (SMC) \citep{gordon_quantitative_2003}. The dust extinction law is sensitive to the properties of the dust, such as its size distribution and chemical composition \citep[e.g.,][]{weingartner_dust_2001, draine_interstellar_2003, hou_evolution_2017}, whereas the dust attenuation law additionally considers the effects of the star-dust geometry on top of those from the dust properties. In simulations, we can see that if we fix the extinction law, the resultant attenuation law varies significantly depending on the dust content and the star-dust geometry \citep[e.g.,][]{seon_radiative_2016, trayford_fade_2020, narayanan_theory_2018, tacchella_h_2022}.

Early work studying the difference between the dust attenuation of nebular emission lines and that of the stellar continuum motivated the idea that different regions in galaxies have different types and amounts of dust \citep[e.g.,][]{fanelli_spectral_1988, calzetti_dust_1994}. These results led to the two-component dust model from \cite{charlot_simple_2000}, where the young stars ($<10$ Myr) experience additional attenuation compared to the older stars. The first dust component is the diffuse dust in the interstellar medium (ISM), which attenuates the light from all stars. The second dust component is the dust in the birth clouds around young stars ($<10$ Myr), which attenuates the light from those young stars and the emission lines from the nebular regions around those young stars. These birth clouds have an approximate dispersal time of $10$ Myr \citep{blitz_origin_1980}, after which the light from these stars will only be attenuated by the ISM dust. However, recent studies find these birth clouds to have a range in typical lifetimes of $10-30$ Myr \citep{chevance_lifecycle_2020}. 

Some studies show that the dust attenuation of nebular regions is not significantly offset from that of the stellar continuum at $z\sim2$ \citep[e.g.,][]{reddy_characteristic_2012, shivaei_investigating_2015, reddy_mosdef_2015}. Therefore, the difference between stellar and nebular dust attenuation is not fixed and has been found to correlate with the specific star formation rate (sSFR) \citep{price_direct_2014, puglisi_dust_2016} and is consistent with the two-component dust model. Galaxies with high sSFR have their young stars dominating the SED, and so the majority of the stellar light and the emission lines from HII regions see the birth cloud dust and ISM dust; therefore, we expect little deviation between the amount of attenuation of the stellar continuum and the emission lines. For galaxies with lower sSFR, the older stars, which only see the ISM dust, contribute significantly to the SED; hence, there will be a greater difference between the attenuation of the stellar continuum and the emission lines. The gas-phase metallicity has also been observed to drive variation between the dust attenuation of the stellar continuum and emission lines \citep{shivaei_mosdef_2020}.

How the dust is modelled in a galaxy can have profound effects on the resultant SED fit of the galaxies, leading to biases in galaxy properties such as the stellar mass and redshift \citep[e.g.,][]{kriek_dust_2013, hahn_inhomogeneous_2024}. At high redshifts, in particular when only rest-frame UV observations are available, there is a strong degeneracy between dust, age and metallicity \citep[e.g.,][]{papovich_stellar_2001}, and the use of a flexible attenuation law allows for this degeneracy to be better understood \citep[e.g.,][]{salim_dust_2018, tacchella_fast_2022, tacchella_stellar_2022}.

Dust attenuation additionally affects the observed sizes of galaxies. The size of a galaxy, or its effective radius, is measured using its light profile. Depending on where the light originates, the measured size will vary relative to the underlying stellar mass distribution, with many works measuring a large divergence between the size of galaxies represented by their light profiles ($R_{\rm light}$) compared to their mass profiles ($R_{\rm mass}$) \citep[e.g.,][]{szomoru_stellar_2013, suess_half-mass_2019, mosleh_galaxy_2020}. 

The ratio of the half-light radius to the half-mass radius ($R_{\rm light}$/$R_{\rm mass}$) is strongly linked to the dust content in galaxies from \cite{zhang_dust_2023}, where they find this ratio can be as high as $2.5$ using modelled galaxies, far higher than any variations in the observed size caused by the inclination of galaxies. The dust in galaxies can lead to such high $R_{\rm light}$/$R_{\rm mass}$ ratios since there is a gradient in the dust attenuation; the light from the centre of galaxies is typically more attenuated by dust than the light from the outskirts \citep{nelson_spatially_2016, tacchella_dust_2018, shen_high-redshift_2022, matharu_first_2023}. The elevated dust attenuation at the centre of galaxies compared to their outskirts flattens the light profile, pushing the effective radius to larger values, which leads to a more extended $R_{\rm light}$ than $R_{\rm mass}$. This effect is, therefore, less pronounced at longer wavelengths as the effect of dust attenuation becomes weaker; however, even imaging from the JWST F444W filter ($\lambda_{\rm rest}=1.48 \mu \rm m$ at $z=2$) does not directly probe the stellar mass profile \citep{zhang_dust_2023}.

The inclination of galaxies strongly affects the amount of observed dust attenuation in local galaxies \citep{zuckerman_reproducing_2021, maheson_unravelling_2024}. However, at cosmic noon, there appears to be a much weaker relationship between dust attenuation and inclination from recent works \citep{lorenz_updated_2023}. They argue that this is due to a clumpy star-dust geometry, where most of the dust attenuation is from large spherical star-forming clumps, which will not be strongly affected by the inclination. These clumps are more common at cosmic noon than in local galaxies \citep{forster_schreiber_constraints_2011, wuyts_smoother_2012}, and so this dust model could explain the evolution in the dependence of the dust attenuation on inclination. Therefore, we do not anticipate the inclination of our galaxies to strongly influence the observed dust attenuation. 

Measurements of dust attenuation laws at higher redshifts towards cosmic noon are challenging, and those that have been measured around cosmic noon have a large diversity. For example, \cite{reddy_mosdef_2020} measure an attenuation curve at $z\sim2$ that is similar in shape to the Milky Way extinction curve, \cite{shivaei_mosdef_2020} measure an attenuation curve at $z\sim2$ that varies with the gas-phase metallicity; for low-metallicity galaxies, the attenuation curve is similar to that of the SMC extinction curve, and for high metallicity galaxies the attenuation curve agrees with the \cite{calzetti_dust_2000} attenuation law. Recently, \cite{sanders_aurora_2024} measured an attenuation law at $z=4.41$ significantly steeper than the \cite{calzetti_dust_2000} law. The exact shape of dust attenuation laws at cosmic noon and how they relate to the properties of galaxies is still an open question. 

In this work, we aim to measure the attenuation laws of a sample of galaxies at cosmic noon to assess how the stellar populations and the morphology of the galaxies relate to dust attenuation. We use spectroscopy from JWST NIRSpec and photometry from NIRCam from the Blue Jay survey, a JWST Cycle 1 program, for a mass-selected sample of galaxies. We obtain $R\simeq1000$ spectra covering $3700$~\AA{}-$1.2~\mu$m rest-frame for all our galaxies, covering the full range of optical emission lines, along with photometry from $3300$~\AA{} to $1.6~\mu$m. We use a Bayesian SED inference code, \texttt{Prospector}, to infer non-parametric SFHs and dust attenuation laws. We measure morphological parameters of our galaxies to see how they are affected by dust attenuation, then compare this dust attenuation of the stellar continuum to that of the nebular regions inferred from the Balmer decrement.

This paper is split into eight sections, with an overview of the data sample in Section~\ref{s:data_sample}, the methods we use to determine physical properties of our galaxies in Section~\ref{s:methods}, the resultant dust attenuation laws and how these compare to other laws from the literature in Section~\ref{s:dust_law_props}. We show the resultant correlations between the dust attenuation law and the stellar population parameters in Section~\ref{s:scaling law}, the relationship between dust attenuation and morphology in Section~\ref{s:morphology}, and the comparison between stellar continuum and nebular dust attenuation in Section~\ref{s:BD_atten}. We conclude our results in Section~\ref{s:conclusion}. Throughout this paper, a cosmology with $H_0=70$ km s$^{-1}$ Mpc$^{-1}$, $\Omega_{\Lambda}=0.7$, and $\Omega_{m}=0.3$ is adopted.

\section{Blue Jay: Data and Sample} \label{s:data_sample}

We use a sample of 153 galaxies from the JWST Cycle 1 program Blue Jay (GO 1810; PI Belli). These galaxies are spread over two pointings in the COSMOS field and were observed with the NIRSpec micro-shutter assembly (MSA; \citealt{ferruit_near-infrared_2022, rawle_-flight_2022}) to obtain spectra with R$\simeq1000$ and a rest-frame coverage of $3700$~\AA{}-$1.2~\mu$m. Four of these galaxies are filler targets at z $\sim$ 6, and data extraction failed for eight galaxies, leaving 141 galaxies forming a mass-selected sample ( $9<\log(\rm M_{\star}/M_{\odot})<11.5$) at cosmic noon ($1.7 < z < 3.5$). The three medium-resolution gratings of NIRSpec (G140M, G235M and G395M) were used to observe all the galaxies in the sample with exposure times of 13h, 3.2h and 1.6h, respectively. A slitlet of at least two MSA shutters was placed on each target, and an A-B nodding pattern was employed along the slitlet. A modified version of the JWST Science Calibration Pipeline v1.10.1 and the Calibration Reference Data System version 1093 are used to reduce the observed spectra for analysis. We derive a master sky spectrum from empty shutters for background subtraction. 

We additionally have parallel deep NIRCam photometry for part of our sample using the F090W, F115W, F150W, F200W, F277W, F356W, F410W and F444W filters. For the galaxies in our sample with NIRSpec measurements, which were offset from our NIRCam field-of-view, public NIRCam imaging from the JWST Cycle 1 program PRIMER (GO 1837; PI Dunlop) is used. Publicly available HST/ACS+WFC3 \citep{skelton_3d-hst_2014, momcheva_3d-hst_2016} photometry, in the F125W, F140W, F160W, F606W and F814W are also used. For further details on the sample selection, observations and data reduction, see the Blue Jay survey paper (Belli et al., in prep).  

The observations of these galaxies have been fruitful, with several papers published on their results. The quenching of star formation is shown to be due to multiphase gas outflows for a massive galaxy in \cite{belli_massive_2023}. Observations of widespread AGN-driven neutral gas outflows at $z\sim2$ are presented in \cite{davies_jwst_2024}, AGN feedback in quiescent galaxies is traced using ionised gas emission in \cite{bugiani_agn_2024}, and widespread rapid quenching is explored in \cite{park_widespread_2024}.

\section{Methods}\label{s:methods}

In this section, we present how we extract the properties of the stellar population, dust attenuation and morphology for the galaxies in this survey. We use SED fitting to infer the stellar mass, SFR, SFH and the dust attenuation law. We discuss the properties of these attenuation laws in Section~\ref{s:dust_law_props} and connect the attenuation law with the stellar population properties in Section~\ref{s:scaling law}. We use NIRCam imaging with multiple bands from the rest-frame optical to rest-NIR to infer the sizes and axis ratios of the galaxies, and use these measurements to assess the relationship between galaxy morphology and dust attenuation in Section~\ref{s:morphology}. We then measure the Hydrogen recombination line fluxes to obtain the Balmer decrement, which is sensitive to the nebular dust attenuation, and we compare this with the measure of the stellar continuum dust attenuation from the \texttt{Prospector} fits in Section~\ref{s:BD_atten}.

\subsection{Stellar Population Modelling}

We model the stellar continuum using \texttt{Prospector}, a Bayesian stellar population inference code \citep{johnson_stellar_2021}. In this work, we build on the \texttt{Prospector} results as presented in \cite{park_widespread_2024}, where the spectra and photometry are fit simultaneously. We adopt the synthetic stellar population library FSPS \citep{conroy_propagation_2009, conroy_propagation_2010}, the MIST isochrones \citep{choi_mesa_2016}, the C3K spectral library \citep{cargile_minesweeper_2020} and assume the Chabrier initial mass function \citep{chabrier_galactic_2003}. We model the star formation history (SFH) using the non-parametric method from \cite{leja_how_2019, leja_older_2019}, with SFH bins logarithmically spaced in age. For further details on the construction of the SFH model, see \cite{park_widespread_2024}. We extract the stellar mass (M$_{\star}$) and SFR from the median posterior of each galaxy. The SFR is measured over the last $30$ Myr.

\texttt{Prospector} accounts for dust absorption and re-emission, which are assumed to be in energy balance. The dust absorption model follows the two-component dust model, with a primary dust component which applies to all stars and follows the \cite{kriek_dust_2013} attenuation curve, as well as a second dust component which only attenuates young stars ($<10$ Myr) which follows a power law attenuation curve. In the following section, we discuss how this is implemented and how the stellar population influences the overall dust attenuation law.

\subsection{Dust Modelling with FSPS}\label{s:dust_modelling}

The attenuation law for the birth cloud dust around young stars ($<10$ Myr) is modelled by a power law in FSPS of the form:

\begin{align}
    \tau_{\lambda, \rm BC} = \ \tau_1 \left(\frac{\lambda}{\lambda_{\rm V}}\right)^{n_1},
\end{align}
where $\tau_{\lambda, \rm BC}$ is the birth cloud optical depth (proportional to the birth cloud dust attenuation law $A_{\lambda, \rm BC}$), $n_1$ is the slope, $\lambda_{\rm V}$ is the wavelength of the visible band at $5500$\AA{} and $\tau_1$ is the optical depth of this dust component at $\lambda_{\rm V}$. We do not include the 2175~\AA\ UV-bump in this attenuation law, since we do not expect bump-forming carbonaceous dust grains, e.g., polycyclic aromatic hydrocarbons  \citep[PAHs;][]{papoular_polycrystalline_2009, steglich_electronic_2010, lin_polycyclic_2025}, to survive in birth clouds around young stars due to their strong UV radiation fields \citep{gordon_dust_1997, witstok_carbonaceous_2023}. In this work, we fix the slope of the birth cloud dust attenuation law to $n_1 = -1$.

The ISM dust attenuation law is modelled following the parameterisation by \cite{noll_analysis_2009} used in \cite{kriek_dust_2013}, following the equation:
\begin{align}
    \tau_{\lambda, \rm ISM} = & \ \frac{\tau_2}{4.05}\left(k(\lambda) + D(\lambda)\right)\left(\frac{\lambda}{\lambda_{\rm V}}\right)^{n_2}, \\
    D(\lambda) = & \ \frac{E_b(\lambda \Delta \lambda)^2}{(\lambda^2 - \lambda_0^2)^2+(\lambda\Delta\lambda)^2}, \\
    E_b = & \ 0.85 - 1.9 \ n_2 \label{eq:bump_conroy}
\end{align}
where $\tau_{\lambda, \rm ISM}$ is the optical depth of the ISM dust (proportional to the ISM dust attenuation law $A_{\lambda, \rm ISM}$), $k(\lambda)$ is the \cite{calzetti_dust_2000} attenuation curve and $D(\lambda)$ is the Lorentzian-like Drude profile to parameterise the 2175~\AA\ UV-bump. $\Delta\lambda$ is is the FWHM of the bump, taken as $350$\AA{} as measured by \cite{noll_analysis_2009}, and $\lambda_0$ is the central wavelength of the bump at $2155$\AA{}. The strength of the bump here, $E_b$, is related to the slope of the attenuation law $n_2$ following the relation above, measured in \cite{kriek_dust_2013}. The only free parameters for the ISM dust are the optical depth, $\tau_2$, and the slope of the attenuation law, $n_2$. 

The overall attenuation law we will see for a galaxy depends on the shape of each attenuation law and its SFH, i.e. how many old and young stars are present. This overall dust attenuation law can be expressed as follows:
\begin{align}
    \label{eq:convolution}
    \tau_\lambda = & \ f_L(\lambda | t<10{\rm Myr}) \cdot [\tau_{\lambda, \rm BC}(\tau_1, n_1) + \tau_{\lambda, \rm ISM}(\tau_2, n_2)]  \\
    & + f_L(\lambda | t>10{\rm Myr}) \cdot \tau_{\lambda, \rm ISM}(\tau_2, n_2) \nonumber
\end{align}
where $\tau_\lambda$ is the overall optical depth (proportional to the overall attenuation law $A_{\lambda}$), $f_L(\lambda | t<10{\rm Myr})$ is the ratio of flux from stars younger than $10$ Myr to the total stellar population, and $f_L(\lambda | t>10{\rm Myr})$ is that for the older stellar population. Here, the SFH informs us of the luminosity of stars older or younger than $10$ Myr, which sets $f_L(\lambda | t<10{\rm Myr})$ and $f_L(\lambda | t>10{\rm Myr})$ and hence controls how much the attenuation law of each dust component affects the overall attenuation law.

These young and old stellar populations have different spectra, with the young stars emitting mainly in the UV and the older stars emitting mainly at redder wavelengths. In Appendix~\ref{App:fl10}, we show that when young ($<10$ Myr) stars make up more than $5\%$ of the stellar mass, they emit over half of the total galaxy flux between $1500-5500$~\AA\ rest-frame. 

To better demonstrate how each dust parameter and the SFH affect the final attenuation law, we create toy model galaxies in FSPS to vary each parameter and determine the resultant attenuation law. We parameterise the SFH through f$_{10}$, the mass ratio of young ($<10$ Myr) stars to the total stellar population. We vary the parameters: $\tau_1$, $\tau_2$, $n_2$, and f$_{10}$ across a grid. We set a fiducial value for each parameter of $\tau_1=0.2$, $\tau_2=0.2$, $n_2=-0.7$ and f$_{10} = 0.5$. As one parameter varies, all other parameters are fixed to their fiducial value, whilst we fix $n_1 = -1$ throughout this section. See Appendix~\ref{App:fiducial} for the specific SFH used to build these models and how we extracted the dust attenuation law from the model.

\begin{figure*}
\centerline{\includegraphics[width=1.0\textwidth]{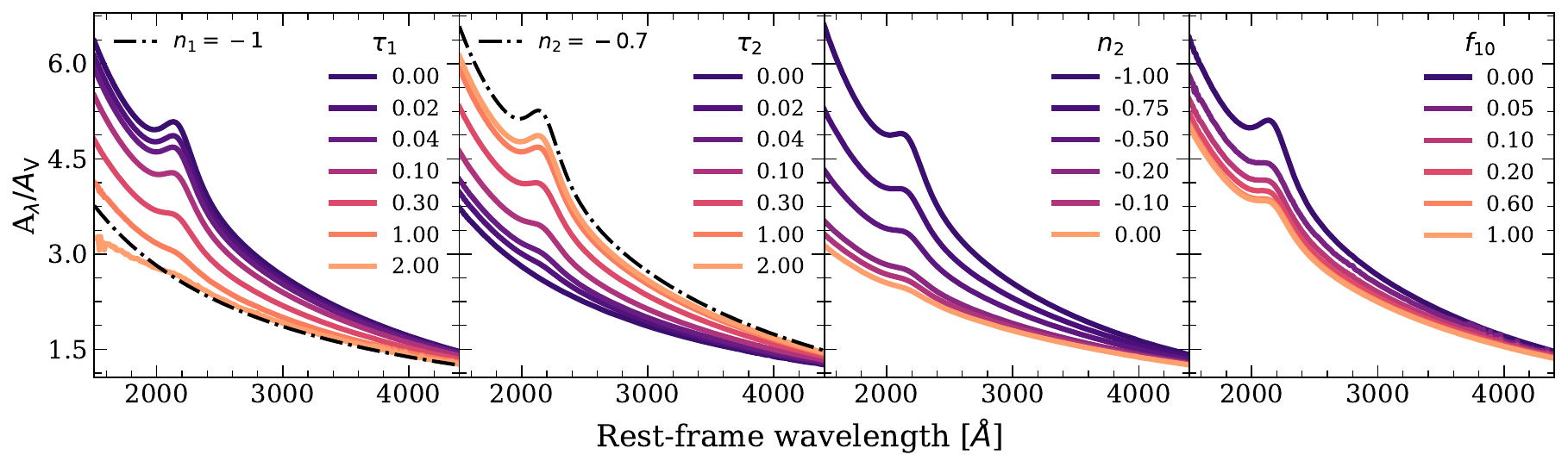}}
\medskip
\caption{The effective attenuation law of a toy model galaxy varying (left to right) the birth cloud optical depth, $\tau_1$; the ISM optical depth, $\tau_2$; the ISM dust attenuation law slope, $n_2$, and the mass fraction of young stars ($<10$ Myr), $f_{10}$. As one parameter varies, all other parameters remain fixed to the fiducial values. In the first panel, as $\tau_1$ increases, the slope of the overall attenuation law decreases towards the slope of the birth cloud dust attenuation law, shown as the black dashed-dot line. In the second panel, the same occurs but for the ISM dust; as $\tau_2$ increases, the slope and bump strength increase towards the ISM dust attenuation law, shown as the black dashed-dot line. In the third panel, as the slope of the ISM dust attenuation law increases, both the bump strength and slope increase due to their relationship in this model. In the last panel, as $f_{10}$ increases, the slope becomes shallower, and the bump strength decreases due to the increased contribution of the birth cloud dust to the overall attenuation law.} \label{f:fsps_dust1}
\end{figure*}

The first parameter varied is $\tau_1$, the optical depth of the birth cloud dust, as shown in the first panel of Figure~\ref{f:fsps_dust1}. When $\tau_1=0$, there is essentially no birth cloud dust; only the ISM dust attenuates the light. As $\tau_1$ increases, we see that the slope of the curve decreases, drawing down towards the birth cloud-only attenuation curve (black dash-dot curve), and the bump strength becomes less substantial as the birth cloud dust (which has no bump) contributes more significantly towards the overall attenuation law. Similarly, $\tau_2$ is varied in the second panel of Figure~\ref{f:fsps_dust1}. When $\tau_2=0$, there is no ISM dust, and we see the attenuation law of the birth cloud dust. The slope of the curve and strength of the UV-bump increase as $\tau_2$ increases since the ISM dust contributes more towards the overall attenuation law. The slope of the ISM attenuation law, $n_2$, is varied in the third panel of Figure~\ref{f:fsps_dust1}. As $n_2$ moves to more negative values, the slope becomes steeper, and the bump strength increases since the slope and bump strength are tied together for the ISM dust law. As $n_2$ approaches zero, the slope of the ISM attenuation law flattens, and the overall attenuation law approaches the slope of the birth cloud dust law, and the bump strength becomes weaker. The final parameter varied is the mass ratio of young ($<10$ Myr) stars, $f_{10}$, shown in the fourth panel of Figure~\ref{f:fsps_dust1}. Here, the slope decreases as $f_{10}$ increases since the stellar population is composed of younger stars, and so more stellar light is being attenuated by the birth cloud dust, which has a shallower slope. Since the birth cloud and the ISM dust attenuation laws are unchanged as $f_{10}$ varies, only their relative contribution to the overall attenuation law is affected. Therefore, the properties of the stellar population and the properties of the dust are essential ingredients in determining the overall attenuation law. 

\subsubsection{Determining the Attenuation Law}

\begin{figure*}
\centerline{\includegraphics[width=0.8\textwidth]{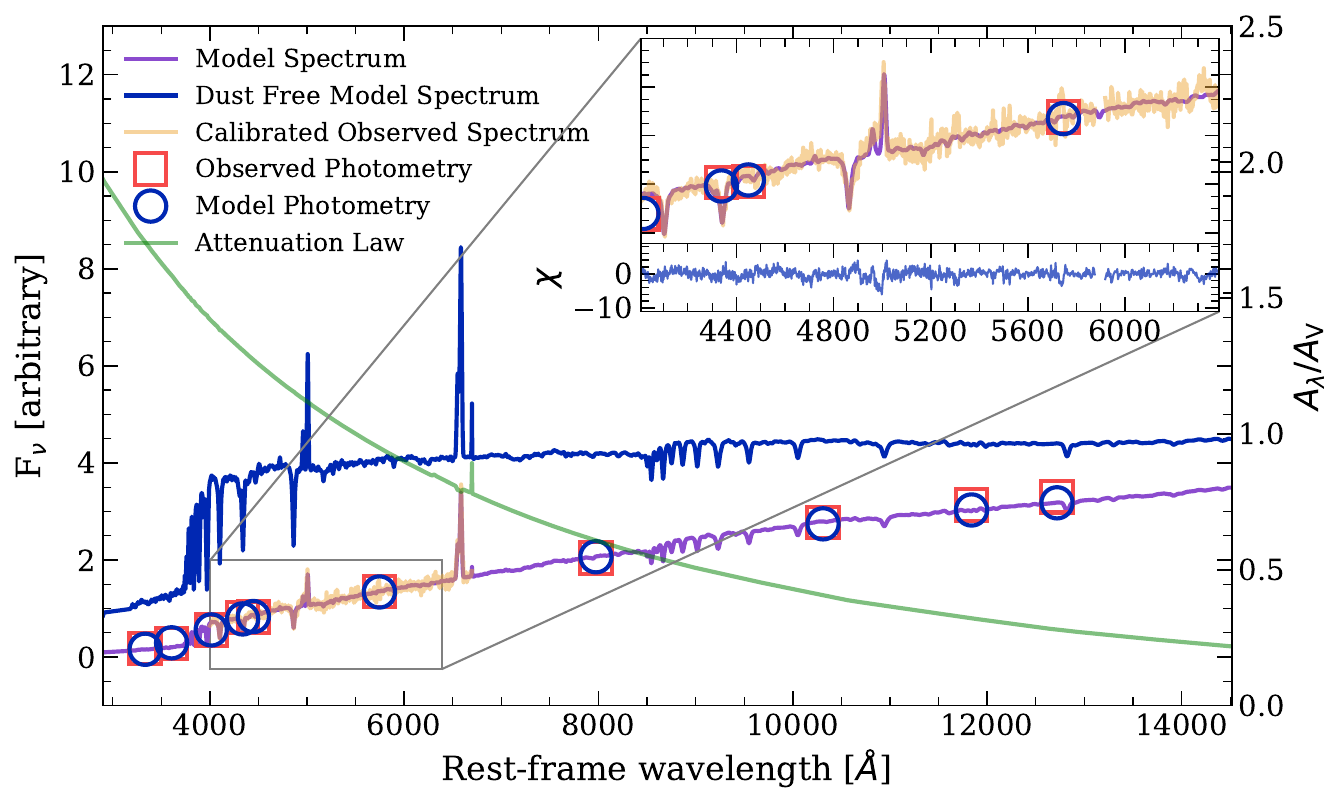}}
\medskip
\caption{The observed and best-fit spectra and photometry for the galaxy with ID$ =11142$. The observed spectrum is fit over the range $4000<\lambda_{\rm rest}<6700$ \AA\, and the residual of this fit is shown in the inset figure. The observed and model photometry are in good agreement. The dust-free spectrum is shown in blue, and the attenuation law for this galaxy is shown in green.} \label{f:prosp_11142}
\end{figure*}

We measure the overall dust attenuation law using the model spectra from the \texttt{Prospector} fit. To demonstrate the high S/N of the data we use in this work, as well as how the \texttt{Prospector} fit is performed, we plot the observed and model photometry and spectra for an example galaxy, COSMOS-11142, in Figure~\ref{f:prosp_11142}. The observed spectrum (calibrated using the photometry) from NIRSpec MSA used in the \texttt{Prospector} fits is plotted as the orange line in both the main figure and the zoomed-in inset. Only spectra between $4000-6700$ \AA\ in rest-frame wavelength are fit for the galaxies in this sample since this includes many of the spectral features which are sensitive to the galaxies' age, such as the Balmer lines, which aid in constraining the SFH, whilst also avoiding the fitting of the $4000$ \AA\ break and the Balmer break, which may affect the flux calibration. We refer the reader to \cite{park_widespread_2024} for further details on how this flux calibration was performed. 

The purple curve represents the best-fit model spectrum, and the blue curve is the model spectrum with no dust. The dust attenuation law was derived by comparing the best-fit spectrum with and without dust, plotted as the green curve. The red squares represent the observed NIRCam photometry, and the blue circles represent the best-fit model photometry. The inset figure shows a zoom-in of the region where the observed spectrum is used in the fitting, along with the residual, $\chi$, between the observed and modelled spectra, with $\chi = $\ (observed-model)/(observed error). This galaxy, COSMOS-11142, is a massive galaxy at $z=2.44$ undergoing rapid quenching, and \citet{belli_star_2024} presents evidence of a powerful neutral gas outflow driven by a supermassive black hole at its centre, which is sufficient to quench the star formation in this galaxy.

\subsection{Emission Line Measurements}

In this work, we use the emission line fluxes calculated in \cite{bugiani_agn_2024}, where the best-fit stellar continuum spectrum is subtracted from the observed spectrum, and the emission lines are individually modelled using Gaussian profiles, with the velocity dispersion and redshift being identical for all lines. Several emission lines are fit, including the H$\alpha\ \lambda6565$, and H$\beta\ \lambda4863$ recombination lines. When determining the velocity dispersion, the contribution to the overall broadening of the lines from the wavelength-dependent spectral resolution of NIRSpec is considered. The Python library \texttt{EMCEE}, an Affine Invariant Markov Chain Monte Carlo (MCMC) ensemble sample \citep{foreman-mackey_emcee_2013}, is used to fit the emission lines. When fitting, the only free parameters are the redshift, velocity dispersion and line flux (except for the lines with fixed or constrained flux ratios). 

We use the Balmer emission lines to measure the Balmer decrement, the ratio of the H$_{\alpha}$ to H$_{\beta}$ emission lines, which traces the dust attenuation of emission lines from nebular regions around young stars. Assuming Case B recombination in these nebular regions (and electron temperature $\rm T=10^4\rm\ K$ and electron density $n_e = 10^2~\mathrm{cm}^{-3}$, see \cite{chakraborty_x-ray_2021} for the applicability of this assumption), the intrinsic ratio of the Balmer decrement is 2.86. Since the H$_{\alpha}$ line is redder than the H$_{\beta}$ line, as dust attenuation increases, the Balmer decrement increases above the intrinsic value. In Section~\ref{s:BD_atten}, we compare the dust attenuation of the light from nebular regions to the stellar continuum dust attenuation measured from \texttt{Prospector} to see how these two dust attenuation tracers relate to each other. 

The galaxies with signal-to-noise (S/N) $~>25$ for H$_{\alpha}$ and S/N$~>3$ for H$_{\beta}$ are selected for this work, as these S/N cuts ensure the Balmer decrement would be measured accurately. The severe cut on H$_{\alpha}$ is equivalent to a cut in SFR, leading to the sample being mainly composed of star-forming galaxies, and ensures we have meaningful detections of the H$_{\beta}$ line, as is done in e.g., \cite{mannucci_fundamental_2010, hayden-pawson_klever_2022, maheson_unravelling_2024}. The small S/N cut on H$_{\beta}$ ensures that it is detected so that the Balmer decrement is reliable. Any higher S/N cut on H$_{\beta}$ would bias the sample to less dusty galaxies, since dustier galaxies attenuate the H$_{\beta}$ line strongly and tend to have weaker detections. This selection criterion leaves 67 galaxies with reliable Balmer decrement measurements.

\subsection{Morphological Measurements} \label{s:morph_measure}

To infer the key morphological parameters of the galaxies in our sample, such as their sizes and axis ratios, we use the fully Bayesian code \texttt{Pysersic} \citep{pasha_pysersic_2023}. We fit each galaxy with a one-component S\'ersic profile, which models its surface brightness and is given by:

\begin{align}
    I(r) = I_e \exp \left[ -b_n \left( \left( \frac{r}{r_e} \right)^{1/n} - 1 \right) \right]
\end{align}
where $I(r)$ is the radial intensity profile at a radius $r$ from the galaxy's centre, $I_e$ is the intensity at the half light radius, $r_e$, and $b_n$ is a function of the S\'ersic index $n$. \texttt{Pysersic} allows us to obtain constraints on the free model parameters: the position angle $\theta$, ellipticity $e$, S\'ersic index $n$, light centroid $(x_0,y_0)$, and half-light radius $r_e$, for each galaxy. The axis ratio b/a is defined as $e=1-~ {\rm b/a}$.

Using \texttt{Photutils} \citep{bradley_astropyphotutils_2024}, source selection and masking are performed before each fit. We create a segmentation map of the central galaxy based on a signal-to-noise cut set to 3, which varies for more intricate systems. We mask all other detected objects outside this map through the fitting procedure. This masking ensures that the derived best-fit parameters are not influenced by flux from other sources. We apply the same source masking to each filter cutout.

To infer the posterior distributions of our free parameters, we use the variational inference technique implemented in \texttt{Pysersic}, which trains a normalising flow to estimate the posterior and then samples directly from it. This offers a speed increase with respect to traditional Markov Chain Monte Carlo (MCMC) sampling while still being able to capture non-Gaussian distributions.

We apply this procedure to the F150W, F356W and F444W imaging, with point spread functions (PSFs) derived following the methodology of \cite{ji_jades_2023}, where WebbPSF models \citep{perrin_updated_2014} are input into JWST level-2 images and are then mosaiced using the same exposure pattern as the NIRCam observations.

We discard the fits with strong residuals as determined through visual inspection. These typically include complex morphological systems, such as mergers, for which the simple one-component model does not provide a good approximation of the flux distribution, as well as AGN and low S/N objects. We obtain a final sample of 104 galaxies selected based on the quality of the fits in the F356W and F444W filters and 98 in the F150W filter. The results of these fits and their relationship to the dust in these galaxies are explored in Section~\ref{s:morphology}.

\subsection{Sample Selection}

From the full Blue Jay sample of galaxies with \texttt{Prospector} fits (141 galaxies), we select three sub-samples of galaxies for analysis in this work. We first use all the galaxies with \texttt{Prospector} fits where we can extract the overall attenuation law in Section~\ref{s:dust_law_props} and Section~\ref{s:scaling law}, leaving  137 galaxies. In Section~\ref{s:morphology} we analyse only the star-forming galaxies ($\log (\rm sSFR ~\left[yr^{-1}\right])>-10$) in our sample that have morphological fits, leaving 93
of the same galaxies in F356W and F444W, and 87 of those in F150W. In Section~\ref{s:BD_atten} we use only the galaxies with sufficient signal-to-noise on the H$_{\alpha}$ (S/N$>25$) and H$_{\beta}$ (S/N$>3$) lines, leaving 67 galaxies.

\section{Properties of the Dust Attenuation Laws}\label{s:dust_law_props}

Investigating how the properties of the dust attenuation laws relate to each other can help us understand how the attenuation law is constructed, and when variations in it are driven by the modelling assumptions or by galaxy properties. In this section, we analyse the shape of the measured attenuation laws for the full sample of galaxies with \texttt{Prospector} fits (137 galaxies). First, we compare the attenuation laws of subsets of galaxies in bins determined by specific galaxy and dust parameters to see overall trends. Then, we investigate how the parameters of the dust attenuation laws relate to each other to understand how flexible these attenuation laws are. We compare these results to those from several extinction and attenuation laws from the literature: the Milky Way extinction law, the LMC and SMC extinction laws, and the \citet{calzetti_dust_2000} attenuation law.

In order to quantify the shape of the overall attenuation law, we measure the value of the attenuation law at $5500$~\AA\ to determine the optical, or visible, dust attenuation ($A_{\rm V}$), then compare this with the value of the attenuaion law in the UV band ($A_{\rm UV}$ at $1500$~\AA) to obtain the UV-optical slope ($A_{\rm UV}/A_{\rm V}$). We then parameterise the strength of the UV-bump at $2175$~\AA\ for the overall attenuation law following the parameterisation from \cite{salim_dust_2020}:

\begin{gather}
    \label{eq:bump}
    B \equiv A_{\rm bump}/A_{2175}  \\
    A_{\rm bump} = A_{2175} - A_{2175, 0}  \\
    A_{2175, 0} = 0.33A_{1500} + 0.67A_{3000}  
\end{gather}
where $B$ is the strength of the bump, $A_{\rm bump}$ is the \textit{additional} attenuation caused by the bump over the rest of the attenuation law, $A_{2175}$ is the overall attenuation at 2175 \AA, and $A_{2175, 0}$ is the attenuation expected at $2175$ \AA{} by using the measured attenuation at $1500$ \AA{} and $3000$ \AA. The relation to calculate $A_{2175, 0}$ was measured in \cite{salim_dust_2020} using simulated attenuation curves from \cite{narayanan_theory_2018}. 

This parameterisation of the bump strength differs from that used in the model of the ISM attenuation law that we adopt when performing the SED fitting (Equation~\ref{eq:bump_conroy}). The method presented here allows us to measure the overall bump strength of the effective attenuation law, whereas the bump strength from Equation~\ref{eq:bump_conroy} only applies for galaxies with no birth cloud dust, where the overall attenuation law is effectively the ISM attenuation law. Therefore, the bump strength we measure for our overall attenuation laws will not necessarily follow Equation~\ref{eq:bump_conroy}. How this impacts the relationship between the measured bump strength and the slope of the overall attenuation law is explored in Section~\ref{s:bump_properties}.

\subsection{Attenuation Law Shape with $A_{\rm V}$, stellar mass and SFR}\label{s:atten_law_shape}

\begin{figure*}
\centerline{\includegraphics[width=1.0\textwidth]{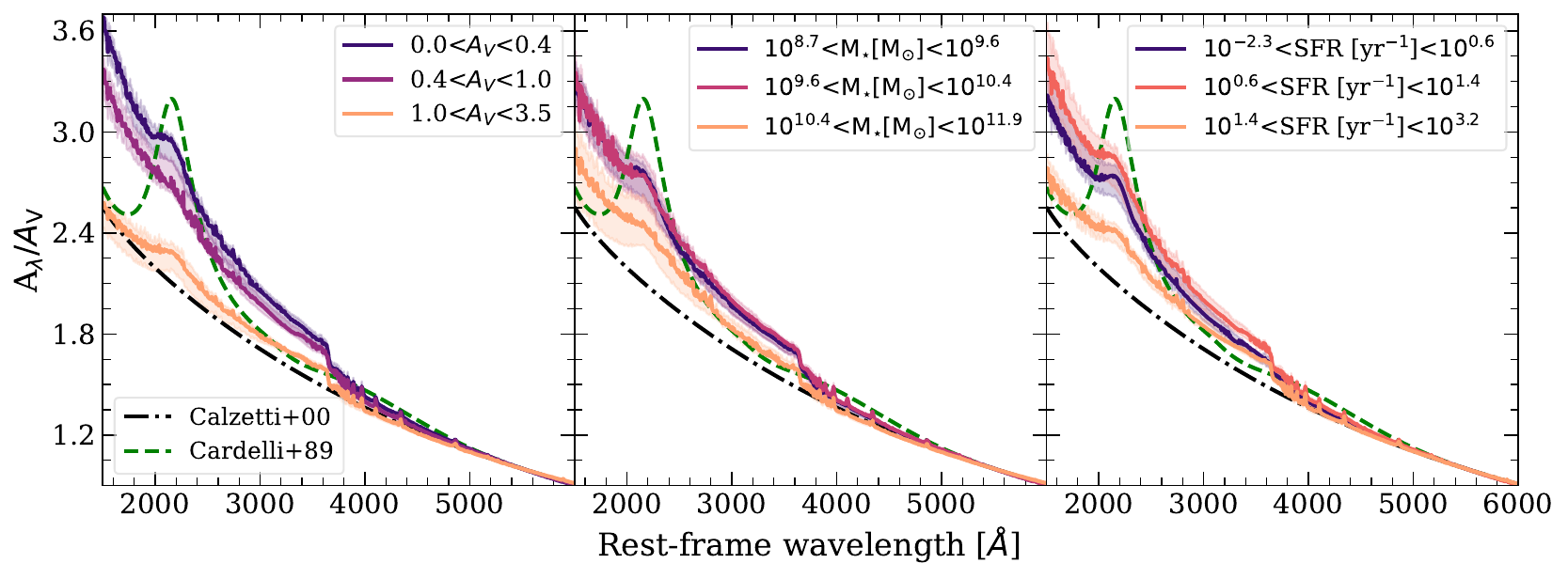}}
\medskip
\caption{The attenuation law, normalised by $A_{\rm V}$, for our sample of galaxies in bins of $A_{\rm V}$, stellar mass ($\rm M_{\star}$) and star formation rate (SFR), from left to right. Dispersion of the attenuation laws is shown as the shaded regions. The attenuation law of local starbursts \citep{calzetti_dust_2000} and the Milky Way extinction law \citep{cardelli_relationship_1989} are plotted as the black dashed-dot and green dashed lines, respectively. Large variation in slope is seen when binning in $A_{\rm V}$, with smaller but significant variation with stellar mass and SFR for $\lambda_{\rm rest}\lesssim3500$\AA. The attenuation law slope becomes flatter for the galaxies with the highest $A_{\rm V}$, stellar mass and SFR. The bump strength has negligible variation and is much weaker than that of the Milky Way.} \label{f:salim_atten}
\end{figure*}

To see how the shape of the attenuation laws vary with different parameters of our galaxies, we first bin the galaxies based on their attenuation in the optical band ($A_{\rm V}$), their stellar mass and their SFR, as shown in Figure~\ref{f:salim_atten}. We performed bootstrap random sampling within each parameter bin using the median attenuation law of each galaxy to determine the 16-84$^{\rm th}$ percentile error ranges of the median attenuation laws. Incorporating the measurement uncertainty of the attenuation law for each galaxy would increase the error on the median attenuation law in each parameter bin. Hence, the errors shown underestimate the true variation within each bin. The attenuation law of local starbursts \citep[black dash-dot line]{calzetti_dust_2000} and the Milky Way extinction law \citep[green dashed line]{cardelli_relationship_1989} are overplotted on each panel for comparison. 

In each panel, the slope of the attenuation laws has little variation for $\lambda_{\rm rest}\lesssim3500$~\AA, however we see more diversity at $\lambda_{\rm rest}\gtrsim3500$~\AA. The largest variation in the slope of the attenuation laws can be seen as we vary $A_{\rm V}$, with the slope becoming shallower as $A_{\rm V}$ increases. The relationship between the slope of the attenuation law and the amount of dust attenuation is discussed further in the next section. Considering the stellar mass and the SFR, the slope of the attenuation law is shallowest for the most massive and star-forming galaxies. 

The strength of the $2175$ \AA\ UV-bump of the median attenuation laws has little variation with $A_{\rm V}$, stellar mass or SFR, and is much weaker than the strength of the bump as seen in the MW \citep{cardelli_relationship_1989}. This is caused by how we model the bump strength, and is discussed in more detail in Section~\ref{s:bump_properties}. 

The overall diversity we observe in these attenuation laws at this epoch demonstrates that assuming a fixed-shape dust law in SED fitting will not accurately represent the dust attenuation, and may therefore bias measurements from the SED fitting, e.g., the stellar mass \citep{hahn_inhomogeneous_2024}.

Although we identify net trends between the shape of the attenuation law and the stellar population properties, it can be more informative to assess these trends when we use the parameters of the attenuation law. This is explored further in Section~\ref{s:scaling law}.

\subsection{UV-Optical Slope of the Attenuation Laws}

A relationship between the optical depth in the visible band ($A_{\rm V}$) and the slope of the attenuation law has been observed in several works \citep[e.g.,][]{salmon_breaking_2016, salim_dust_2018, nagaraj_bayesian_2022}. They find that as $A_{\rm V}$ increases, the slope of the attenuation law decreases and becomes shallower. This trend has also been investigated using theoretical attenuation laws \citep{chevallard_insights_2013} and further supported with hydrodynamic simulations \citep{narayanan_theory_2018, trayford_fade_2020}. 

\begin{figure}
\centerline{\includegraphics[width=.5\textwidth]{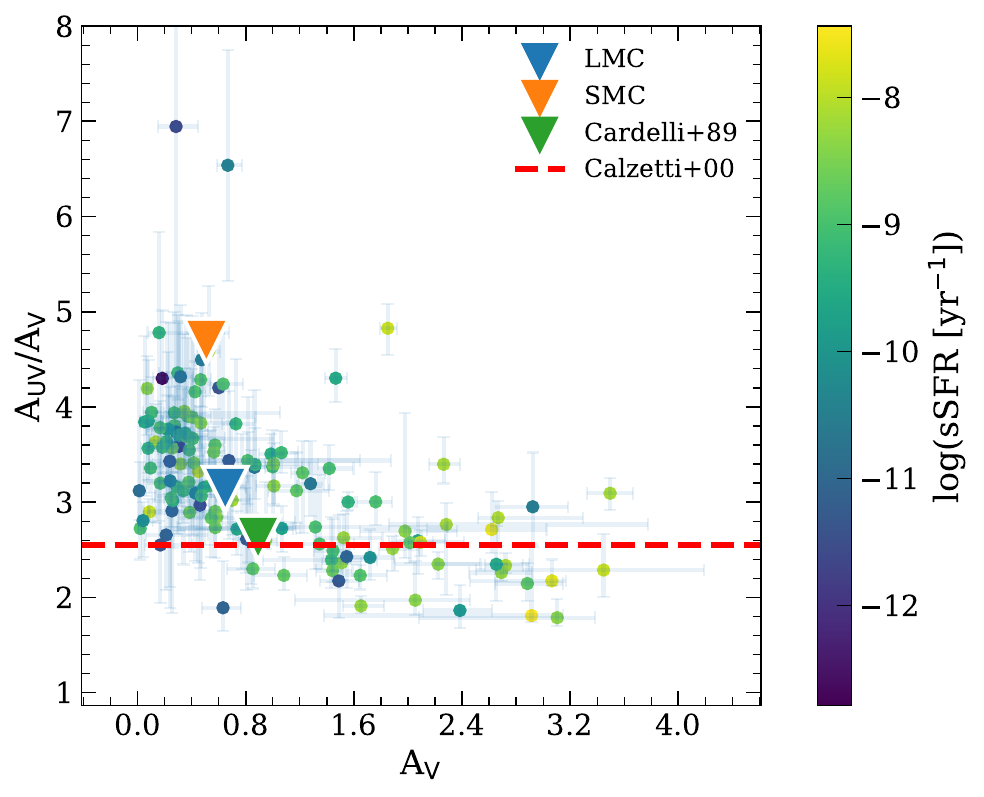}}
\medskip
\caption{The UV-optical slope of the attenuation law ($A_{\rm UV}/A_{\rm V}$) plotted against the dust attenuation in the optical ($A_{\rm V}$) for our sample of galaxies colour-coded by the sSFR. The MW \citep{cardelli_relationship_1989}, LMC and SMC \citep{gordon_quantitative_2003} dust laws are overplotted as inverted triangles. The \citealt{calzetti_dust_2000} dust attenuation law is plotted as a horizontal red dashed line since it has a fixed slope but can have a variable $A_{\rm V}$, and our measured attenuation laws can reproduce these literature values. No galaxies have both high $A_{\rm V}$ and a steep slope, and these galaxies are on average the most star-forming in the sample. For galaxies with low dust attenuation, there is a large spread in the value of the UV-optical slope.} \label{f:salim_slope_Av}
\end{figure}

We find a similar trend for the attenuation laws in this work. We show how the optical attenuation, $A_{\rm V}$ relates to the UV-optical slope, $A_{\rm UV}/A_{\rm V}$, in Figure~\ref{f:salim_slope_Av}. We overplot, as inverted triangles, the values of the optical attenuation and the UV-optical slope for the LMC, SMC and the MW extinction curves, and as a dashed red line for the variable \cite{calzetti_dust_2000} attenuation law. From this plot, we can see that the effective attenuation laws we fit from \texttt{Prospector} can reproduce the slope and optical attenuation values of the dust laws from the literature. A clear trend is apparent, with galaxies having a large range in slopes for low $A_{\rm V}$, and as $A_{\rm V}$ increases, the slopes of the attenuation laws flatten.

Although $A_{\rm V}$ is on both axes, this trend can be physically explained by the fact that red photons scatter more isotropically than blue in the optical \citep{chevallard_insights_2013}. 
For a face-on galaxy, as photons are emitted into the equatorial plane, the red photons would have a higher chance of being scattered out of the plane, and hence out of the galaxy and towards the observer, than the blue photons. These blue photons are then more likely to be absorbed by dust than the red photons, leading to a steeper attenuation curve. This effect is prevalent in the low optical depth limit, steepening the attenuation law. However, when a galaxy has a high dust content, most of the UV radiation from young stars will be attenuated, and any young stars which happen not to be attenuated along our line of sight will dominate the UV regime of the SED. This reduces the observed attenuation in the UV and flattens the slope of the attenuation law. Referring back to Figure~\ref{f:salim_slope_Av}, we can see that the most star-forming galaxies in our sample, which are most likely to have some unobscured young stars, on average have an elevated $A_{\rm V}$ and flatter UV-optical slope than the rest of the galaxies. 

\subsection{2175 \AA{} UV-Bump Strength} \label{s:bump_properties}

The strength of the UV-bump at $2175$ \AA{} has been measured to correlate strongly with the UV-optical slope by \cite{kriek_dust_2013}. In this work, the dust attenuation law for the ISM dust follows this relationship by construction (see Section~\ref{s:dust_modelling}), however, the overall attenuation law combines the ISM dust attenuation law and the birth cloud dust attenuation law, where we model the birth cloud dust law to have no bump. Given how we model the attenuation laws, in this section, we assess the strength of the bump of the overall attenuation law to understand how it may be biased or restricted by these assumptions.

\begin{figure}
\centerline{\includegraphics[width=.5\textwidth]{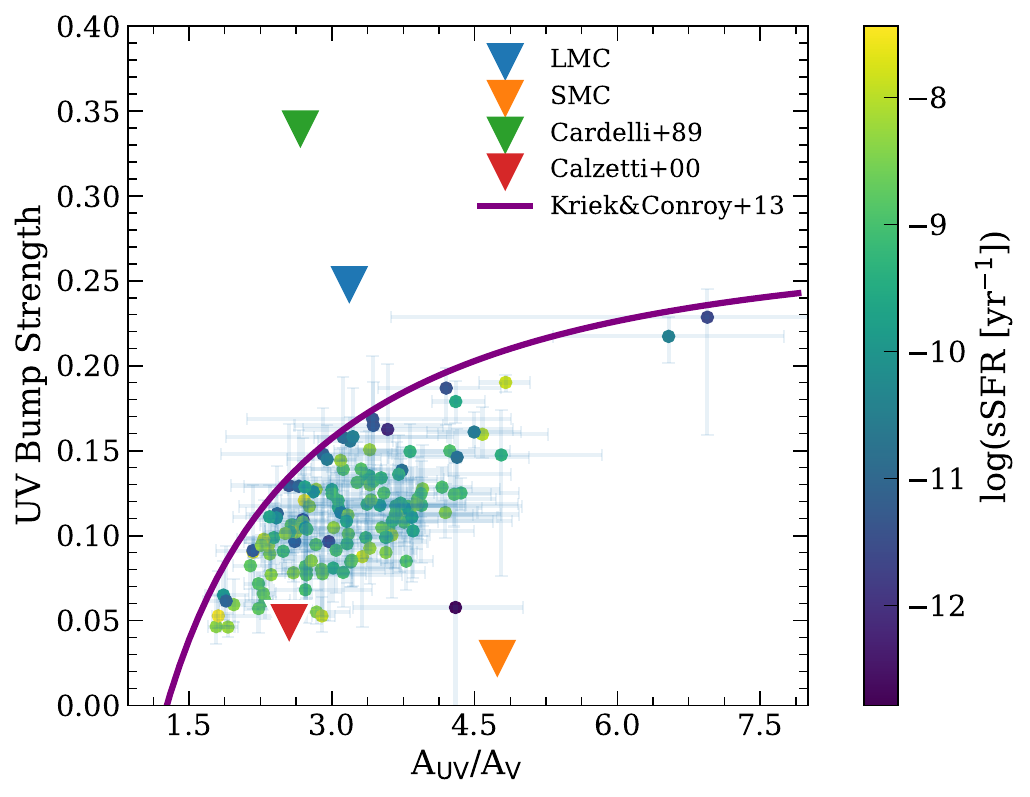}}
\medskip
\caption{The 2175 \AA\ UV-bump strength plotted against the UV-optical slope ($A_{\rm UV}/A_{\rm V}$) of the dust attenuation laws for our galaxies, colour-coded by the sSFR. The values of the slope and bump strength for the MW \citep{cardelli_relationship_1989}, LMC and SMC \citep{gordon_quantitative_2003} and \citealt{calzetti_dust_2000} laws are plotted as inverted triangles, and the \citealt{kriek_dust_2013} relation between these parameters is plotted in purple. More actively star-forming galaxies generally lie further from the purple curve, since the birth cloud dust, which has no bump, significantly contributes to the effective attenuation law. The least star-forming galaxies have low birth cloud dust, and their attenuation laws will approximate the ISM dust attenuation law, which follows the relation between the slope and bump strength shown in the purple curve. Given our model, no galaxy can lie above the purple curve, so we cannot measure attenuation laws with bumps as strong as the MW and LMC extinction laws.} \label{f:salim_bump}
\end{figure}

In Figure~\ref{f:salim_bump}, we plot the UV-bump strength against the UV-optical slope ($A_{\rm UV}/A_{\rm V}$). The strength of the UV-bump and the slope of the dust law for the LMC, SMC, and MW extinction laws and the \cite{calzetti_dust_2000} attenuation law are plotted as inverted triangles. The relationship between the UV-bump strength and the UV-optical slope from \cite{kriek_dust_2013} is plotted as the purple curve, where we apply our parameterisation of the UV-bump strength (Equation~\ref{eq:bump}) to the \cite{kriek_dust_2013} attenuation law at different values of $A_{\rm UV}/A_{\rm V}$. The measured effective attenuation curves rarely reproduce the \cite{kriek_dust_2013} relation since the effective attenuation law is a combination of both the ISM dust attenuation law and the birth cloud dust law. The least star-forming galaxies, which have little birth cloud dust and the overall attenuation law will be dominated by the ISM dust law, lie closest to the \cite{kriek_dust_2013} relation. For the most star-forming galaxies, the birth cloud dust significantly contributes towards the overall attenuation law, so the bump strength is generally reduced further from the purple curve. Therefore, by construction, no galaxy can lie above the purple curve since the ISM dust would put it exactly on the purple curve, and the birth cloud dust will pull it down further to weaker bump strengths. Because of this, our dust attenuation laws are typically unable to reach bump strengths as strong as the MW or the LMC, although we do not expect to have attenuation laws with such high bump strengths at these redshifts \citep[e.g.,][]{shivaei_mosdef_2020}.

Physically, this relation can be understood from the star-dust geometry. Even if a galaxy contains carbonaceous dust grains (e.g., PAHs) that produce a UV-bump, specific star-dust geometries can remove the bump from the observed attenuation law of a galaxy \citep[e.g.,][]{seon_radiative_2016, narayanan_theory_2018}. If any unobscured young stars are present, typical with clumpy star-dust geometries, their UV radiation infills the absorption caused by the carbonaceous dust grains, removing the bump and reducing the UV attenuation, thereby flattening the slope of the attenuation law. If there are no exposed young stars, the bump will be observable, and the slope will be steeper. This simplified model, of course, does not take into account variations in the attenuation law slope or bump strength from the dust grain properties, such as their size distribution and composition \citep[e.g.,][]{hou_evolution_2017}.  

\section{Dust Attenuation Law and Stellar Populations}\label{s:scaling law}

\begin{figure*}
\begin{tabular}{cc}
\includegraphics[width=0.7\textwidth]{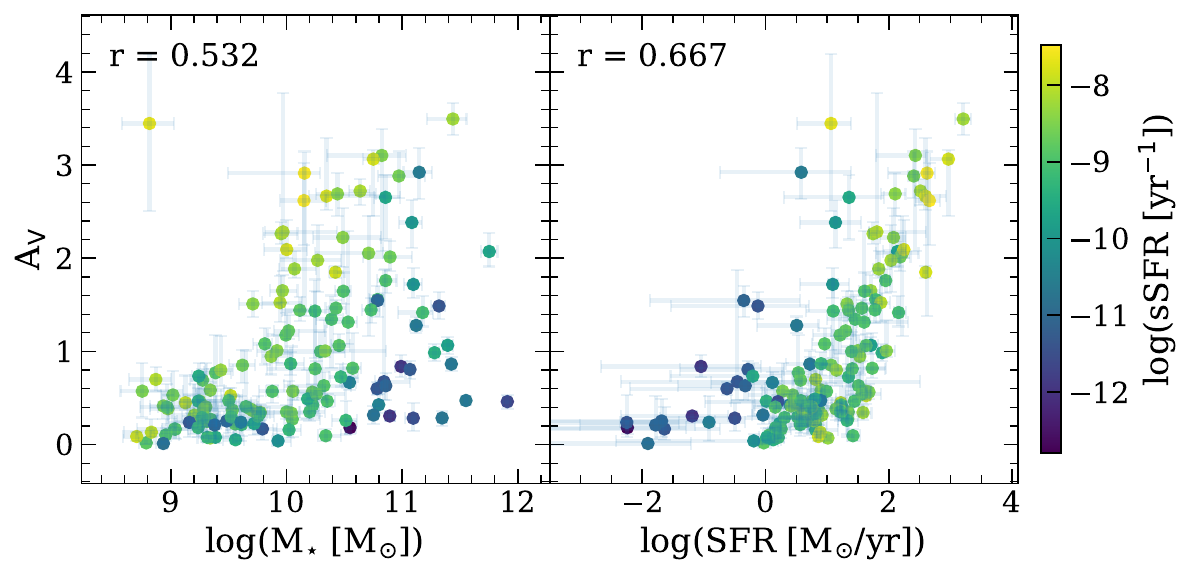} \\
 \includegraphics[width=0.7\textwidth]{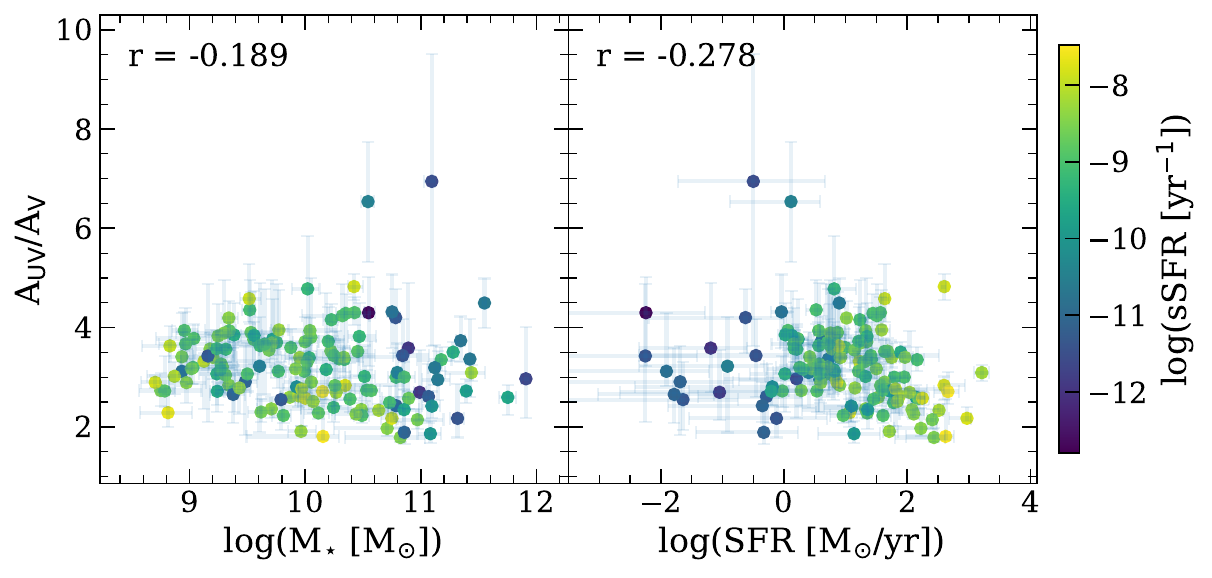} \\
 \includegraphics[width=0.7\textwidth]{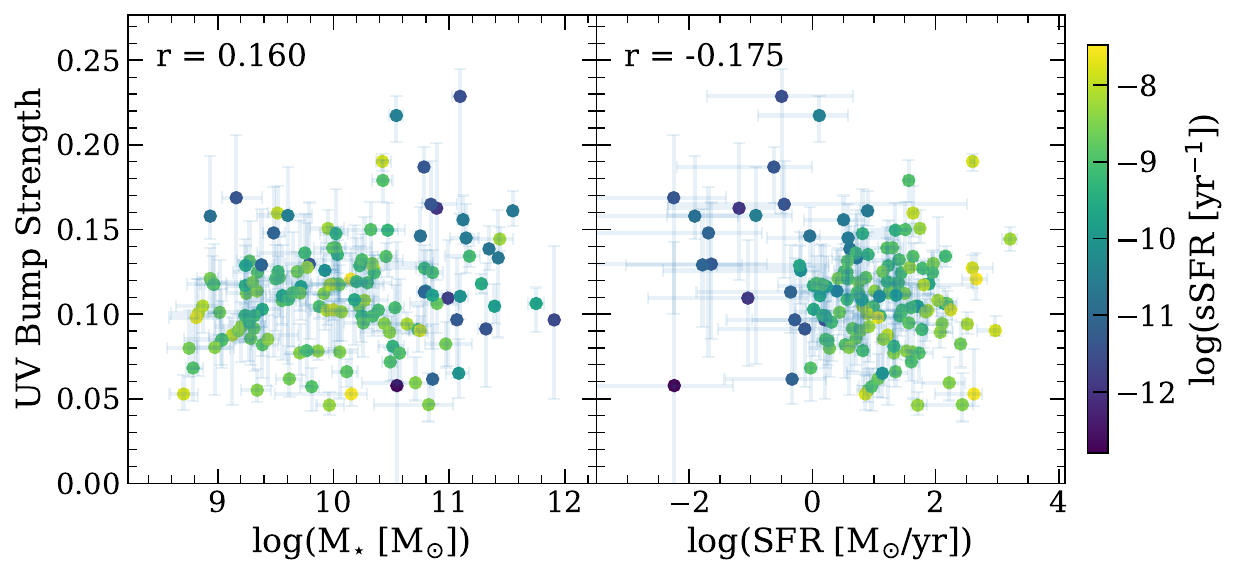} \\
\end{tabular}
\caption{The dust attenuation law parameters plotted against the stellar mass ($\rm M_{\star}$) and star formation rate (SFR), all colour-coded by the sSFR. The Spearman rank coefficient $r$ is given in each panel, quantifying the strength of the correlation between the X and Y parameters. In the top panel, the attenuation in the optical band, $A_{\rm V}$, strongly correlates with the stellar mass and more so with the SFR. Galaxies with larger stellar masses will have built up a larger dust reservoir in their ISM over time, increasing the amount of dust attenuation. Additionally, galaxies with a higher SFR will have more young stars, which are still embedded in their birth clouds, increasing the amount of dust attenuation. These two effects cause the scatter in the dust attenuation versus stellar mass plot. The UV-optical slope ($A_{\rm UV}/A_{\rm V}$) in the middle panel does not correlate with stellar mass; it does, however, show a trend with the SFR. Galaxies with higher SFR tend to have shallower slopes, and galaxies with lower SFR have a large spread in slopes. This is by construction, as the birth cloud dust attenuation law has a shallower slope than the ISM attenuation law, and in highly star-forming galaxies, the birth cloud dust contributes significantly to the overall attenuation law. The bump strength similarly has a weak correlation with stellar mass, and the most star-forming galaxies tend to have a weaker bump strength, again by construction, since we model the birth cloud dust to have no bump strength.} \label{f:Av_mass_SFR}
\end{figure*}

In Section~\ref{s:atten_law_shape}, we saw how the stellar population parameters, namely the stellar mass and SFR, can drive variation in the effective attenuation law. In this section, we investigate how the different parameters of the attenuation laws relate to the stellar mass and SFR using the full sample of galaxies with \texttt{Prospector} fits (137 galaxies). 

We plot the stellar mass and SFR against the parameters of the effective dust attenuation law in Figure~\ref{f:Av_mass_SFR}, namely the dust attenuation in the optical ($A_{\rm V}$), the slope between the optical and the UV regions ($A_{\rm UV}/A_{\rm V}$), and the strength of the $2175$~\AA{} UV-bump. First, looking at the top left plot, we see that $A_{\rm V}$ increases with stellar mass, and this relationship is well-established for star-forming galaxies \citep[e.g.,][]{garn_predicting_2010, shivaei_mosdef_2020, maheson_unravelling_2024, sandles_jades_2024}. As galaxies increase in stellar mass, they build up more dust in their ISM, increasing the amount of dust attenuation. Compared to those works, here we also consider quiescent galaxies, occupying the highest mass and lowest $A_{\rm V}$ region of the plot. It can be seen that the relationship between the mass and the dust attenuation follows tracks depending on the sSFR, i.e., at a given sSFR (or colour), dust attenuation increases as stellar mass increases. This relationship applies even to the least star-forming galaxies in our sample. Although we have few quenched galaxies, the trend is still present, with the Spearman rank correlation between stellar mass and $A_{\rm V}$ being $r=0.64$ for quenched galaxies with $\log(\rm sSFR [yr^{-1}])<-11$ (8 galaxies). Therefore, we are able to measure the stellar mass versus dust attenuation relation for starbursts, star-forming galaxies and quiescent galaxies.

In the top right panel of Figure~\ref{f:Av_mass_SFR} we can see that the SFR is tightly correlated with $A_{\rm V}$, which can be interpreted as the increase in the amount of birth cloud dust attenuating the young stellar light, leading to an increase in the overall dust attenuation. The value of the Spearman rank coefficient between $A_{\rm V}$ and the SFR is $r=0.667$, as shown in Figure~\ref{f:Av_mass_SFR}, which is stronger than the correlation between $A_{\rm V}$ and the stellar mass, $r=0.532$, indicating they both strongly correlate with dust attenuation, but the SFR is slightly more important. The Spearman rank correlation between $A_{\rm V}$ and the sSFR is $0.305$, so comparing the dust attenuation with the stellar mass and SFR is more informative than looking at the sSFR alone.

Considering how the stellar mass and the SFR relate to the dust attenuation, the presence of starbursts, star-forming galaxies, and quenched galaxies in our sample leads to a large variation in dust attenuation values at a given stellar mass. Massive quenched galaxies will have lower dust attenuation than their equal-mass, star-forming counterparts. Similarly, low-mass starbursts will have higher dust attenuation than typical star-forming galaxies at the same stellar mass.

The stellar mass and SFR are, however, not independent quantities, with a plethora of works studying the observed relationship between the SFR and stellar mass, or the star-forming main sequence \citep[e.g.,][]{brinchmann_physical_2004, nelson_where_2016, lin_almaquest_2019, nelson_spatially_2021, leja_new_2022, alsing_pop-cosmos_2024, mcclymont_thesan-zoom_2025}. Therefore, cross-correlations may elevate the correlation between the dust attenuation and the stellar mass or SFR. More advanced statistical analysis taking into account this cross-correlation, as is done in, e.g., \cite{bluck_are_2020}, \cite{baker_almaquest_2022} and \cite{maheson_unravelling_2024}, is not feasible here due to the small sample size. 

Now, considering the UV-optical slope and bump strength of the attenuation laws in the bottom two panels of Figure~\ref{f:Av_mass_SFR}, we can see that they have much weaker correlations with the stellar mass and the SFR compared to $A_{\rm V}$. The slope and bump strength have a negligible correlation with the stellar mass, as is indicated by the small values of the Spearman rank correlations. There is a trend between the slope and bump with the SFR. First, considering the UV-optical slope, the most star-forming galaxies tend to have a shallower slope, and the least star-forming galaxies have a large spread in their slopes. This relationship is by construction, since the birth cloud dust is modelled to have a fixed, shallow slope. When the birth cloud dust contributes significantly to the attenuation law (namely in highly star-forming galaxies), the overall attenuation law will have a shallow slope. The ISM dust is modelled to have a variable slope, so when this dominates the attenuation law in less star-forming galaxies, we see a wider range in the slope of the attenuation law. The same effect is reflected in the relationship between the UV-bump strength and the SFR to a lesser extent, since the birth cloud dust is modelled to have no bump strength.

Another important parameter in galaxy evolution, the stellar metallicity, is expected to have strong correlations with both the stellar mass and the SFR \citep[e.g.,][]{looser_stellar_2024}. The stellar metallicity was, however, poorly constrained for many of the galaxies in our sample due to the lack of absorption lines.

\section{Dust Attenuation Law and Morphology}\label{s:morphology}

Many processes determine how large a galaxy can grow and the shape it settles into, such as mergers \citep{naab_minor_2009, jackson_extremely_2022}, AGN activity \citep{lilly_surface_2016, dubois_horizon-agn_2016} and its SFH \citep{tacchella_evolution_2016,  lilly_surface_2016}. Local star-forming galaxies can be widely described as having disc-like morphologies \citep[e.g.,][]{vincent_dependence_2005, padilla_shapes_2008}; however, at cosmic noon, star-forming galaxies show more diversity in their shapes, with many low-mass galaxies having prolate or ellipsoidal shapes, and the higher mass galaxies having a more disc-like shape \citep[e.g..][]{van_der_wel_geometry_2014, zhang_evolution_2019}.

To assess the morphology of our galaxies and relate this to the dust attenuation, we measure the effective radius ($R_e$) and axis ratio (b/a), as described in Section~\ref{s:morph_measure}. The effective radius, or half-light radius, is the radius within which half of the light emitted is contained, and the axis ratio, b/a, is the ratio of the semi-major and semi-minor axes of a galaxy's projected 2D image on the sky. To go from the axis ratio to an inclination, one would need to assume an intrinsic 3D shape of the galaxy; however, as mentioned above, there is a large diversity in shapes among the star-forming population at these redshifts. It would then be difficult to differentiate between an inclined thin disk and an edge-on thick disk. With our small sample size, it is therefore difficult to accurately measure the inclination of these galaxies. 

In this section, we use the morphological parameters derived from the F150W, F356W, and F444W filters for the star-forming galaxies ($\log(\rm sSFR ~[yr^{-1}])>-10$) within our full sample, leaving morphological measurements for 93 of the same galaxies in F356W and F444W, and 87 of those in F150W. Given our redshift range of $1.7<z<3.5$, the F150W filter probes $0.33<\lambda_{\rm rest}<0.55\mu$m, F356W probes $0.79<\lambda_{\rm rest}<1.32\mu$m and F444W probes $0.97<\lambda_{\rm rest}<1.62\mu$m. We first look at how dust attenuation correlates with the surface density of stellar mass and SFR and how this compares with the global stellar mass and SFR. We then investigate the dependence between dust attenuation and axis ratio recorded at lower redshifts \citep[e.g.,][]{zhang_dust_2023} to see how this extends to our galaxies at higher redshifts. Lastly, we compare the size-mass relations we derive here with other size-mass relations from the literature, and explore the impact dust attenuation has on the measured sizes in different filters. 

\subsection{Dust Attenuation and Stellar Mass and SFR Surface Density}

\begin{figure*}
\begin{tabular}{cc}
  \includegraphics[width=0.8\textwidth]{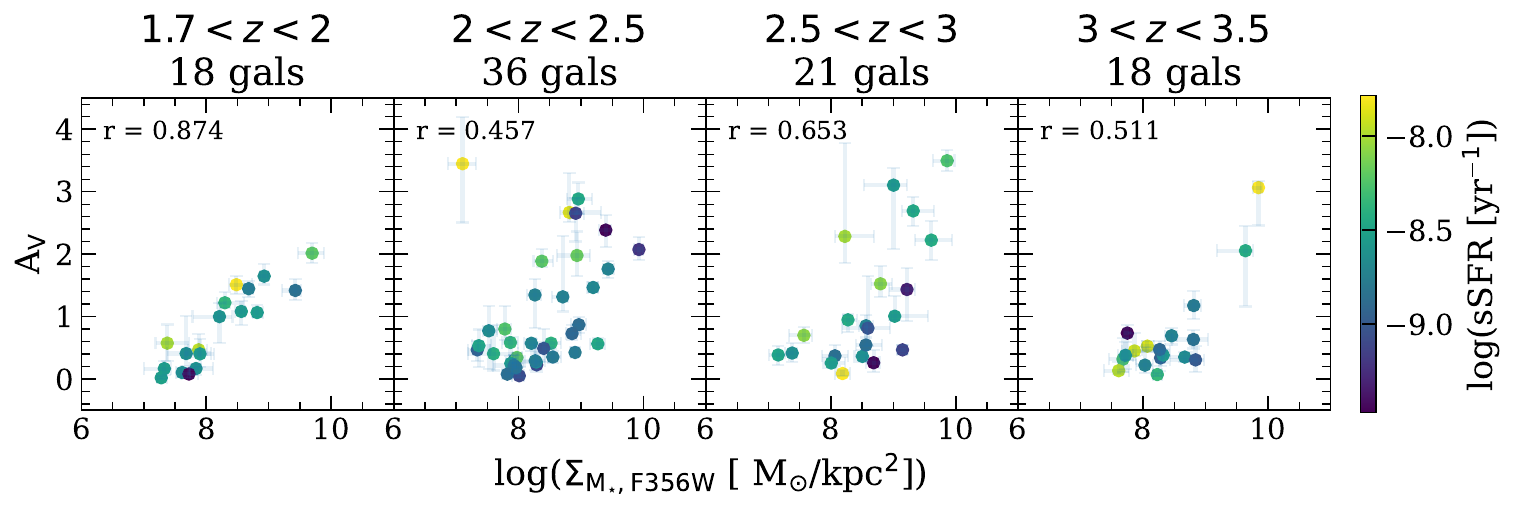} \\
 \includegraphics[width=0.8\textwidth]{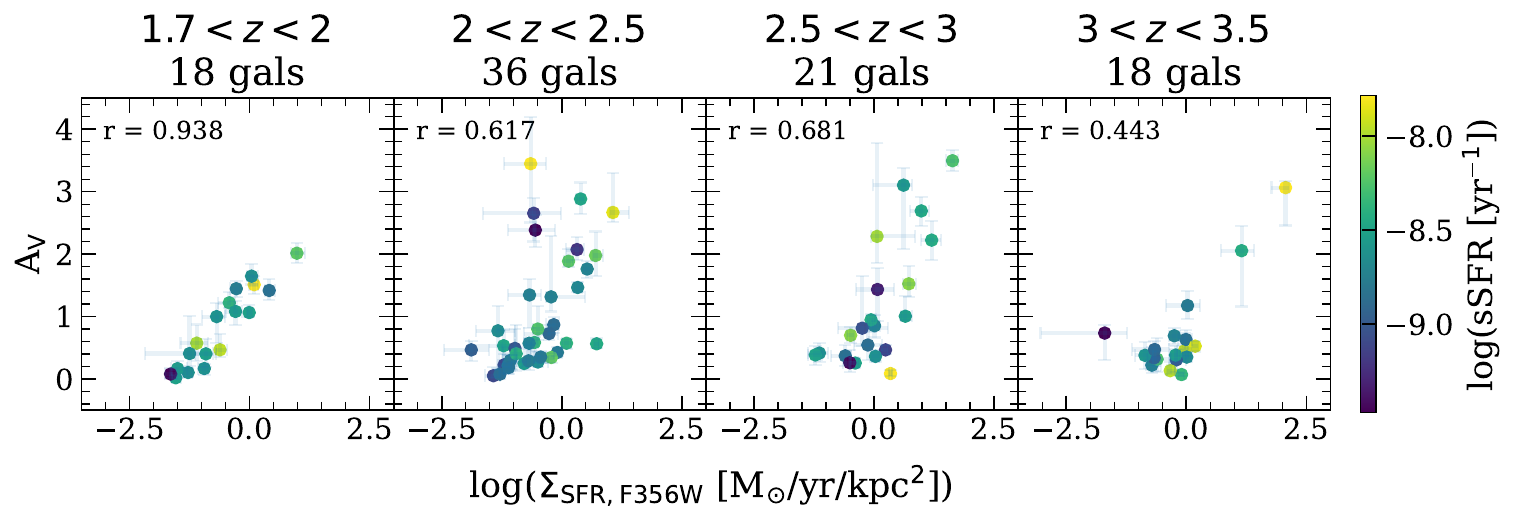} \\
\end{tabular}
\caption{The optical dust attenuation, $A_{\rm V}$, plotted against the surface density of stellar mass, $\Sigma_{\rm M_{\star}}$ in the top panel, and the surface density of SFR, $\Sigma_{\rm SFR}$ in the bottome panel, in bins of redshift, colour coded by the sSFR. The sizes are measured in the F356W filter. The Spearman rank coefficient is written on the figure as 'r', quantifying the strength of the correlation between the X and Y parameters. In both panels, the correlation is tighter than that between $A_{\rm V}$ and the global stellar mass or SFR, shown in Figure~\ref{f:Av_mass_SFR}, implying the surface densities of the stellar mass and SFR are more informative in determining $A_{\rm V}$.} \label{f:Av_surf_density}
\end{figure*}

Measuring the surface density of the stellar mass and SFR tells us how concentrated the stars and star formation are in a galaxy. Since the stellar mass and SFR strongly correlate with the dust attenuation, we expect the surface density of these parameters to also strongly correlate.

In Figure~\ref{f:Av_surf_density}, we show $A_{\rm V}$ plotted against the stellar mass surface density ($\Sigma_{\rm M_{\star}}$) and the SFR surface density ($\Sigma_{\rm SFR}$). We plot the galaxies in bins of redshift to minimise the variation in the rest-frame wavelength at which the sizes are measured. Here, the effective radius we use to determine the resolved parameters is measured using the F356W filter. When using the size measured from the F150W or F444W filters, the rest-frame wavelength probed for each galaxy will be different, and this does lead to strong variations in the measured sizes, as is discussed in Section~\ref{s:size_mass_relations}. When the sizes from different filters are used, the exact value of the SFR and stellar mass surface densities vary; however, the relationship between $A_{\rm V}$ and these parameters does not change significantly. The Spearman rank correlations between $A_{\rm V}$ and the surface density of stellar mass and SFR become slightly larger with filter wavelength, although this effect is small. 

We can see that the relationship between $A_{\rm V}$ and the stellar mass and SFR surface densities are tight, especially in the lowest redshift bin. In most redshift bins, $A_{\rm V}$ relates more strongly to the SFR and stellar mass surface density than the global parameters. The Spearman rank coefficient between $A_{\rm V}$ and the global stellar mass is $0.532$ and with the global SFR is $0.667$. Comparing with the values shown in each panel of Figure~\ref{f:Av_surf_density}, the surface density of the stellar mass and SFR correlate more strongly with $A_{\rm V}$ than the global stellar mass and SFR for the redshift bins $1.7<z<2$ and $2.5<z<3$, with the Spearman rank coefficients being of comparable size in the other redshift bins. However, due to the small sample size in each redshift bin, the global parameters provide a better population-level understanding of how the dust attenuation relates to the stellar population properties.

These results agree with recent work measuring the attenuation law from simulations, where the SFR surface density is more important than the global SFR in determining $A_{\rm V}$ \citep{sommovigo_learning_2025}. Future measurements of the relationship between the dust attenuation and stellar mass and SFR surface densities for larger samples of galaxies will help us better understand whether these surface densities are always more important in determining dust attenuation than the global parameters alone.

\subsection{Dust Attenuation and Axis Ratios}

\begin{figure*}
\begin{tabular}{cc}
  \includegraphics[width=0.8\textwidth]{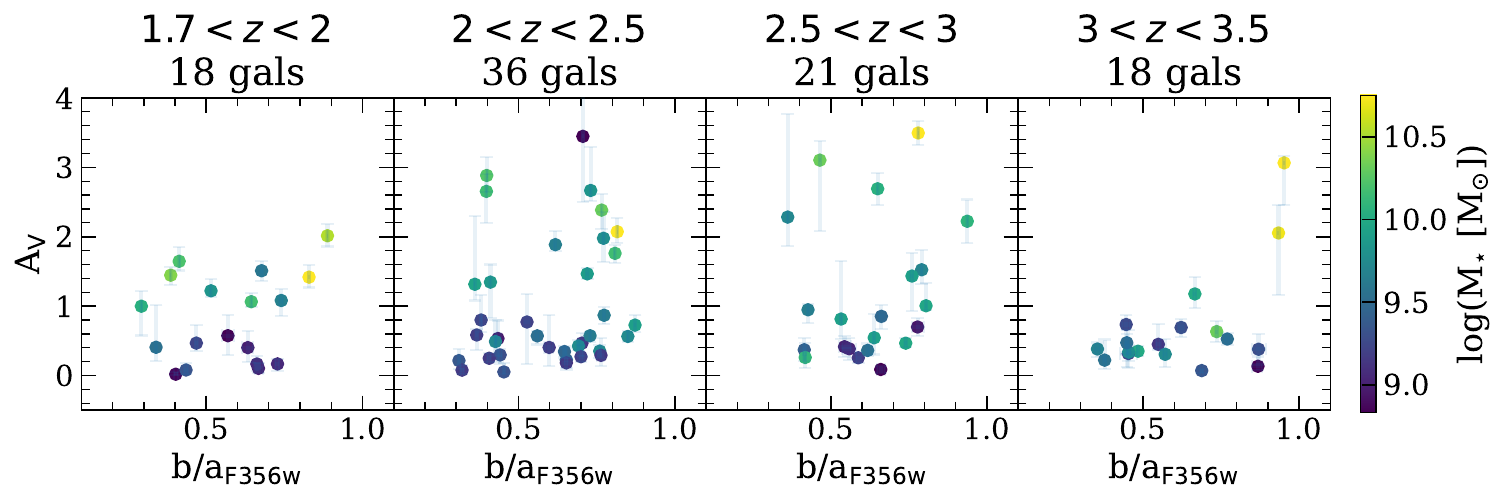} \\
\end{tabular}
\caption{The optical dust attenuation, $A_{\rm V}$, plotted against the axis ratios measured in the F356W filter, b/a$_{\rm F356W}$, in bins of redshift, colour-coded by the stellar mass. There is negligible correlation between $A_{\rm V}$ and the axis ratio except in the highest redshift bin, which is likely driven by the small sample size. The flat relation between $A_{\rm V}$ and the axis ratio is mostly expected at these redshifts, and any variation in $A_{\rm V}$ can be tied to the stellar mass varying, as can be seen by the colour coding.} \label{f:axis_ratio_z}
\end{figure*}

Although the axis ratio is not equivalent to the inclination, we can roughly say that a disc-like galaxy with a low axis ratio will be almost edge-on, and one with a high axis ratio will be almost face-on. Here, we investigate how the axis ratios of our galaxies relate to the dust attenuation, $A_{\rm V}$ in Figure~\ref{f:axis_ratio_z}. We plot the galaxies in bins of redshift to minimise the variation in the rest-frame wavelength at which the axis ratios are measured. 

We can see that in all redshift bins, there is no significant correlation between the dust attenuation and the axis ratio, except in the highest redshift bin, where the galaxies with the highest axis ratio have an increase in dust attenuation. However, due to the small sample size, this effect is likely caused by random scatter and is not physically motivated. Any variation in $A_{\rm V}$ can be seen to be driven by the stellar mass, and not the axis ratio, when we consider the colour-coding; at any axis ratio, as stellar mass increases, $A_{\rm V}$ increases. 

Since we expect a combination of disc-like, ellipsoidal and spheroidal galaxies in the redshift range we are studying in this work, the axis ratio is not strongly related to the inclination for all galaxies. The axis ratio will vary little as the viewing angle changes for an ellipsoidal or spheroidal galaxy. Therefore, the dust attenuation will remain roughly constant with axis ratio, as we observe in Figure~\ref{f:axis_ratio_z} and is demonstrated in \cite{zhang_evolution_2019}. Additionally, a study using observations from HST for star-forming galaxies at $0<z<2.5$ also finds a negligible correlation between dust attenuation and axis ratio for galaxies at $z>1$ \citep{zhang_dust_2023}. Therefore, at cosmic noon, we do not expect to see any strong relationship between the dust attenuation and the axis ratio due to the morphologies of the galaxies. 

\subsection{Size-Mass Relations} \label{s:size_mass_relations}

\begin{figure}
\includegraphics[width=0.5\textwidth]{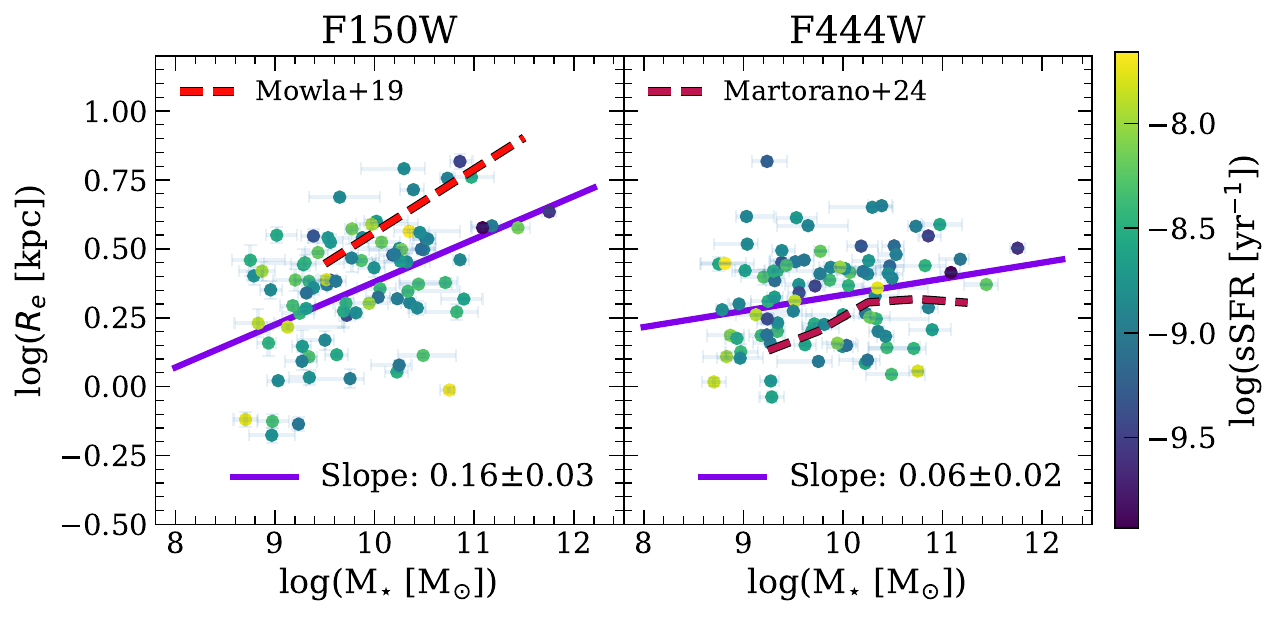}
\medskip
\caption{Size-mass relations for our sample of galaxies in the F150W and F444W filters, colour-coded by the sSFR. The F150W filter covers the rest-frame optical, and the F444W filter covers the rest-frame NIR. Overplotted is the linear best fit of the size-mass relation as the solid purple line in each plot. The rest-optical size-mass relation from \citep{mowla_cosmos-dash_2019} is plotted on the left as the dark orange dashed line, and the rest-NIR size-mass relation from \citep{martorano_sizemass_2024} is plotted on the right as the dark brown dashed line. We measure a positive slope for the size-mass relation in the rest-optical, whereas we measure a nearly flat size-mass relation in the rest-NIR. Our rest-optical size-mass relation is offset to slightly smaller sizes than the \citep{mowla_cosmos-dash_2019} relation, and our rest-NIR size-mass relation is offset to slightly larger sizes than the \citep{martorano_sizemass_2024} relation, on average. Although due to the large scatter of our measured sizes, both size-mass relations measured here are overall consistent with the literature size-mass relations at the same rest-frame wavelengths.} \label{f:size_mass_relations}
\end{figure}

\begin{figure*}
\begin{tabular}{cc}
   \includegraphics[width=0.8\textwidth]{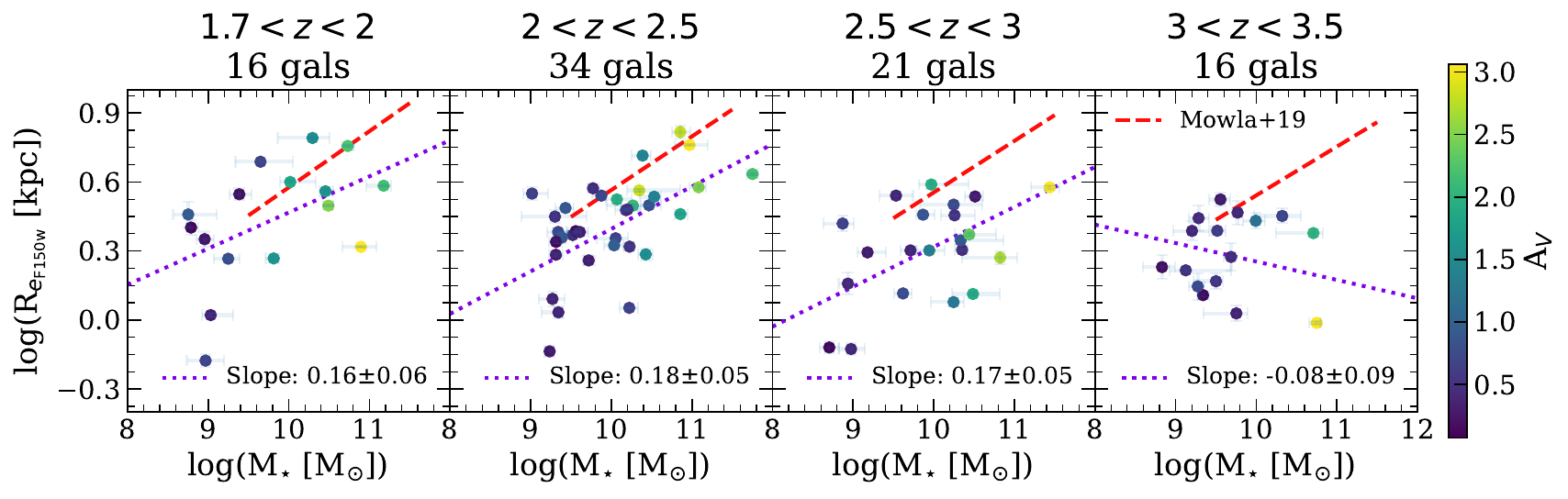} \\
 \includegraphics[width=0.8\textwidth]{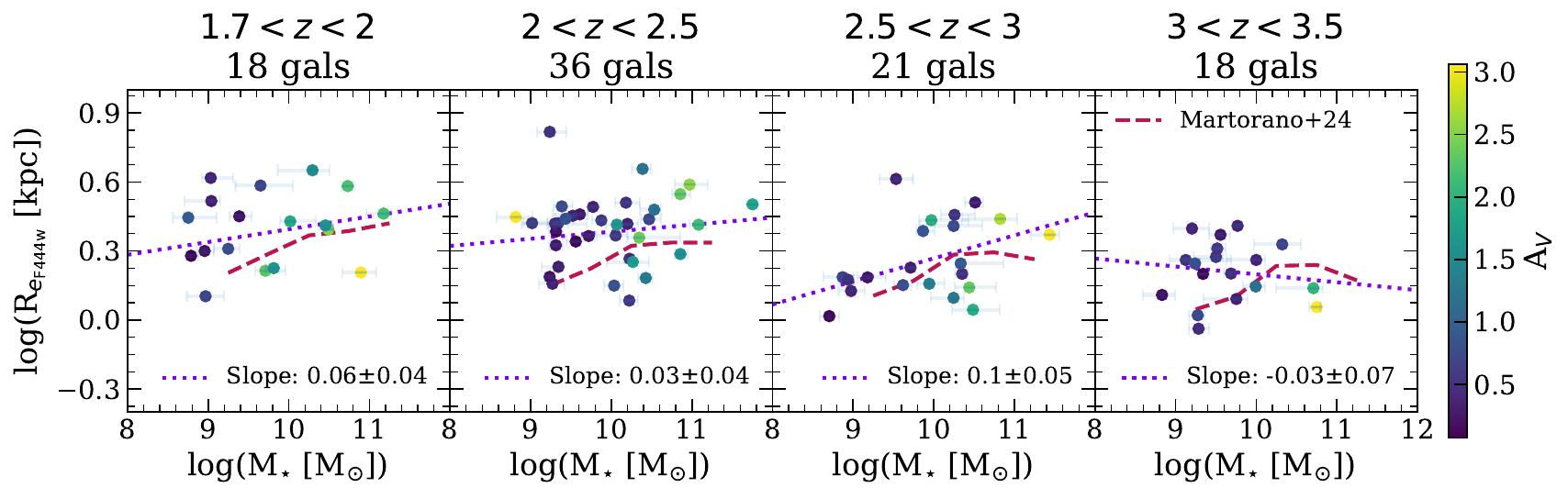} \\
\end{tabular}
\caption{Size-mass relations for our sample of galaxies measured in the rest-optical with the F150W filter (top panel) and measured in the rest-NIR with the F444W filter (bottom panel), in bins of redshift, colour coded by $A_{\rm V}$. The linear best fit is plotted in purple in each panel. The rest-optical size-mass relation from \citealt{mowla_cosmos-dash_2019} for star-forming galaxies is plotted in the top panel as the orange dashed line, and the rest-NIR size-mass relation from \citealt{martorano_sizemass_2024} is plotted in the bottom panel as the brown dashed line. The rest-optical size-mass relation has a positive slope for $1.7<z<3$ but flattens off at higher redshifts, and the rest-NIR size-mass relation is roughly flat across all redshifts. Within each redshift bin, the most massive galaxies (M$_{\star}\gtrsim 10^{10}\rm~ M_{\odot}$) appear to become more extended in the rest-optical than the rest-NIR, due to the elevated dust attenuation of these massive galaxies.} \label{f:size_mass_relations_z}
\end{figure*}

The stellar mass of a galaxy has been recorded to correlate strongly with its size, referred to as the size-mass relation \citep[e.g.,][among others]{cebrian_effect_2014, van_der_wel_geometry_2014, suess_half-mass_2019, mowla_cosmos-dash_2019, yang_evolution_2021, kawinwanichakij_hyper_2021, ward_evolution_2024, allen_galaxy_2024, yang_cosmos-web_2025}. This relation has been measured out to $z=8$ for star-forming galaxies \citep{miller_jwst_2024}, and the slope of this relation is shown to not evolve significantly with redshift. 

In this work, we measure the effective radii of the galaxies in several wide-band filters. Due to our large redshift range ($1.7<z<3.5$), we measure a range in rest-frame wavelength within a single filter. This can lead to strong variations in the measured size-mass relation, since the size has a strong dependence on the rest-frame wavelength at which it is measured \citep[e.g.,][]{ward_evolution_2024}.

To explore how the sizes we measure for our galaxies are affected by the rest-frame wavelength probed, we use the F150W and F444W filters, which provide a large enough baseline for our sample such that there is no overlap in the measured rest-frame wavelengths between the filters. The F150W filter covers $0.33<\lambda_{\rm rest}<0.55\mu$m, or the rest-frame optical, and the F444W filter covers $0.97<\lambda_{\rm rest}<1.62\mu$m, or the rest-frame NIR. In this section, we compare the sizes we measure in the rest-optical with the \cite{mowla_cosmos-dash_2019} size-mass relation, also measured in the rest-optical, and we compare our sizes in the rest-NIR with the \cite{martorano_sizemass_2024} size-mass relation, measured in the rest-NIR. The \cite{mowla_cosmos-dash_2019} relation is measured with the HST/H160 filter up to $z=3$, corresponding with a rest-frame wavelength range of $0.40<\lambda_{\rm rest}<0.64\mu$m for our redshift range of $1.7<z<3$, overlapping well with the rest-frame coverage of F150W used here. The \cite{martorano_sizemass_2024} size-mass relation is measured in the rest-NIR at $\lambda_{\rm rest} = 1.5\mu$m using JWST/NIRCam up to $z=2.5$, which lies within the rest-frame coverage of F444W here.

In Figure~\ref{f:size_mass_relations}, we plot the effective radius ($R_e$) against the stellar mass for the F150W filter, or rest-optical, in the left panel, and the F444W filter, or rest-NIR, in the right panel. We fit a straight line to each size-mass relation to determine its slope, shown as the purple line. We over-plot the rest-optical \cite{mowla_cosmos-dash_2019} size-mass relation for star-forming galaxies in the left panel, and the rest-NIR \cite{martorano_sizemass_2024} size-mass relation for star-forming galaxies in the right panel. Each of these size-mass relations is redshift dependent; therefore, here we use the size-mass relation at the median redshift of our sample ($z=2.46$), and only plot these size-mass relations over the mass ranges in which they were measured. 

Our sizes measured in F150W, or the rest-optical, correlate strongly with the stellar mass, and the slope of this size-mass relation ($0.16 \pm 0.03$) is comparable with that of the \cite{mowla_cosmos-dash_2019} size-mass relation ($0.23$). The sizes in F444W, or the rest-NIR, have negligible correlation with the stellar mass, and the slope of the size-mass relation is almost flat ($0.06 \pm 0.02$), comparable with the slope of the roughly flat size-mass relation from \cite{martorano_sizemass_2024}.

The sizes of our galaxies in the rest-optical appear more compact than the \cite{mowla_cosmos-dash_2019} relation on average. However, the scatter on the \cite{mowla_cosmos-dash_2019} size-mass relation is large enough to cover our measured sizes, and with rest-optical measurements for only 87 galaxies here, it is difficult to make any strong conclusions when comparing these results with other studies.

The flattening of our rest-NIR size-mass relation compared to the rest-optical size-mass relation can be seen to be mainly driven by the most massive galaxies ($\rm M_{\star}\gtrsim10^{10~}M_{\odot}$) being more compact in the rest-NIR than the rest-optical. This colour gradient in the size is explored further in Section~\ref{s:size_grad}.

Since our galaxies cover a relatively wide redshift range ($1.7<z<3.5$), it can be more informative to look at the size-mass relations in redshift bins. In Figure~\ref{f:size_mass_relations_z}, we show the rest-optical size-mass relation from the F150W filter in the top panel, and rest-NIR size-mass relation from the F444W filter in the bottom panel, now colour-coded by $A_{\rm V}$. We plot the linear best-fit in each panel as the purple dashed line. We plot the \cite{mowla_cosmos-dash_2019} rest-optical size-mass relation in the top panel as the orange dashed line, and the \cite{martorano_sizemass_2024} rest-NIR size-mass relation in the bottom panel as the brown dashed line. The rest-optical size mass relation has a strong positive slope for $1.7<z<3$; however, above $z=3$, the most massive, and most dusty, galaxies become more compact, and the size-mass relation flattens. For the rest-NIR, we measure a size-mass relation that is very shallow or consistent with flat in all redshift bins. 

Our rest-optical size mass relation has a similar slope to the rest-optical \cite{mowla_cosmos-dash_2019} relation for $1.7<z<3$. At higher redshifts, we find a much shallower size-mass relation with a slope that is consistent with flat, although the \cite{mowla_cosmos-dash_2019} relation is not measured above $z=3$, and so in our highest redshift bin ($3<z<3.5$), their size-mass relation is just an extrapolation. Additionally, due to the small sample size of galaxies in this redshift bin, we can not draw any strong conclusions from the presence of a flat size-mass relation for $3<z<3.5$.

The size-mass relation we measure in the rest-NIR agrees with the slope of the \cite{martorano_sizemass_2024} relation for all redshifts. However, their size-mass relation was only measured up to $z=2.5$, so we are extrapolating the redshift dependence above this.

When comparing the sizes of the galaxies between the rest-optical and rest-NIR within each redshift bin, it can be seen that the low mass galaxies (M$_{\star}\lesssim 10^{10} ~\rm M_{\odot}$) are at roughly the same sizes, however there is more variation for the high mass galaxies (M$_{\star}\gtrsim 10^{10} ~\rm M_{\odot}$). These galaxies appear more compact in the rest-NIR than in the rest-optical. These galaxies also have the highest dust attenuation, which is likely causing this size colour gradient. This is explored further in the next section.

\subsection{Size Gradients and Dust Attenuation}\label{s:size_grad}
\begin{figure}
\centerline{\includegraphics[width=0.5\textwidth]{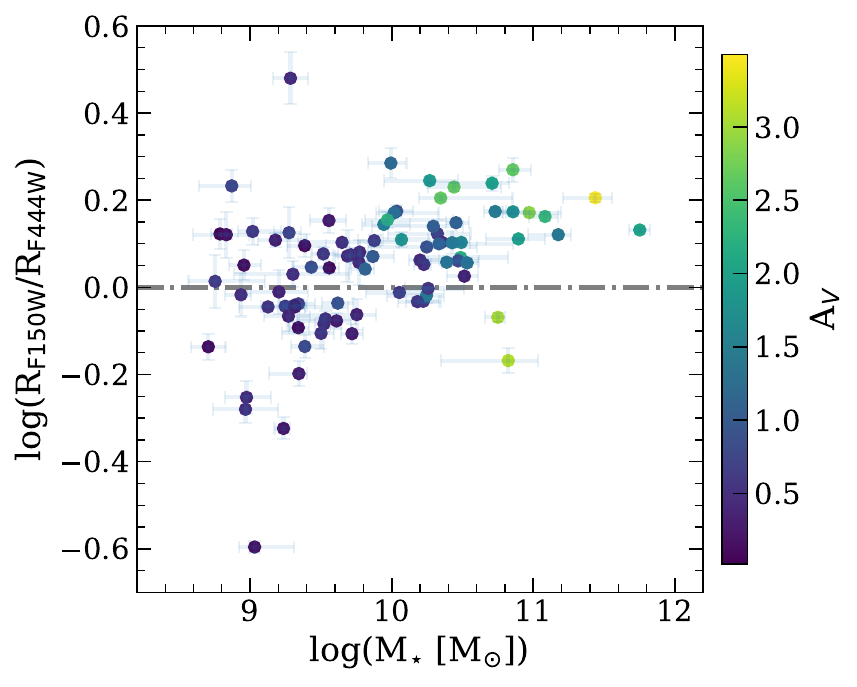}}
\medskip
\caption{Ratio of the effective radius measured in the F150W filter ($\rm R_{F150W}$) to that measured in the F444W filter ($\rm R_{F444W}$), plotted against the stellar mass, colour-coded by $A_{\rm V}$. The F150W filter measures the rest-frame optical here, and the F444W the rest-frame NIR, so $\rm R_{\rm F150W}/R_{\rm F444W}$ is a ratio of the rest-optical to rest-NIR sizes. For the most massive galaxies ($\rm M_{\star}\gtrsim10^{10}~ M_{\odot}$), as $A_{\rm V}$ increases, $\rm R_{\rm F150W}/R_{\rm F444W}$ increases, and the galaxies are more extended in the rest-optical than the rest-NIR, showing the effect of dust attenuation flattening the light profile and extending the effective radius more in the rest-optical than the rest-NIR. For the lower mass galaxies ($\rm M_{\star}\lesssim10^{10}~ M_{\odot}$), there is minimal dust attenuation, so any variation in the size ratios is due to differing galaxy growth pathways, either inside-out growth or a central starburst.} \label{f:size_bias_total}
\end{figure}

To better quantify the effect of dust attenuation on the measured sizes of our galaxies, we use the ratio of the effective radius in the rest-optical, measured in F150W, to the effective radius in the rest-NIR, measured in F444W, ($\rm R_{\rm F150W}/R_{\rm F444W}$). This quantifies if a galaxy appears more extended in the rest-optical or the rest-NIR. In Figure~\ref{f:size_bias_total} we plot this ratio against the stellar mass, colour coded by the amount of dust attenuation $A_{\rm V}$. 

For the most massive galaxies ($\rm M_{\star}\gtrsim10^{10}~M_{\odot}$), a gradient in the size ratios can be seen with the amount of dust attenuation; as galaxies have increased dust attenuation, the sizes become more extended in the rest-optical compared to the rest-NIR. Since dust attenuation is typically higher in the centre of galaxies, the central light will be attenuated more than the light from the outskirts. This leads to the light profile in the rest-optical being flatter and pushing the effective radius to larger values than in the rest-NIR, where the effect of dust attenuation is less. Therefore, as the amount of dust attenuation increases, the size in the rest-optical becomes more extended compared to the size in the rest-NIR. On average, this effect leads to $\log(\rm R_{F150W}/R_{F444W})\sim 0.1$ for the dusty galaxies ($A_{\rm V}>1.5$), corresponding with a size increase of $\sim30\%$ from the rest-NIR to the rest-optical.

Dust attenuation is thus causing the apparent flattening of the size-mass relation in the rest-NIR compared to the rest-optical. A similar effect has been observed from the rest-UV to the rest-optical at $0.5<z<3$, where the slope of the size-mass relation becomes shallower at longer rest-frame wavelengths due to dust attenuation \citep{nedkova_uvcandels_2024}. 

Now considering the lower-mass galaxies ($\rm M_{\star}\lesssim 10^{10}~M_{\odot}$), we instead see a large spread in the size ratio around zero. These galaxies have negligible dust attenuation, so the main driver of this variation will be where the recent star formation has occurred, since younger stars emit more at shorter wavelengths (rest-UV to rest-optical) and older stars emit more at longer wavelengths. If inside-out growth has occurred, where star formation begins at the centre of a galaxy and over time moves outwards, we expect an older population of stars in the centre of that galaxy and a young population of stars on the outskirts \citep[e.g.,][]{munoz-mateos_specific_2007, pezzulli_instantaneous_2015, lian_inside-out_2017, frankel_inside-out_2019, lyu_primer_2025}. This would produce a more extended light profile in shorter wavelengths than longer wavelengths; hence, the rest-optical size will be larger than the rest-NIR size, and $\rm R_{\rm F150W}/R_{\rm F444W}>1$. If, however, a galaxy underwent a central starburst, the centre of that galaxy would be populated with young stars, and the outskirts would have older stars, which could potentially be caused by mergers \cite[e.g.,][]{tacchella_confinement_2016, mcclymont_thesan-zoom_2025-1}. This would lead to the light profile being more extended in the longer wavelengths than the shorter wavelengths, and the size in the rest-optical will be more compact than the size in the rest-NIR, therefore $\rm R_{\rm F150W}/R_{\rm F444W}<1$. 

The effect of these two growth pathways may also be at play in the most massive galaxies in our sample. The effect of dust attenuation leads to the galaxies being more extended in the rest-optical than the rest-NIR; however, we do see some dusty galaxies that are more extended in the rest-NIR than the rest-optical. This could be due to strong inside-out growth or even non-standard dust gradients; however, due to the small sample size, no strong conclusions can be made. 

These results agree with other studies on the effect of the observed filter on the size of galaxies. \cite{suess_rest-frame_2022} report a similar result with the ratio of the sizes in NIRCam F150W and F444W having a dependence on mass, for both star-forming and quenched galaxies in a redshift range of $1.0<z<2.5$, however they do not relate this size gradient to the dust content of the galaxies. From a more theoretical point of view, \cite{popping_dust-continuum_2022} uses simulated galaxies to show that at $z=5$, the size measured at observed-frame $1.6\mu$m can be as much as four times larger than the half-mass radius. Taking this concept to the extreme, \cite{nelson_jwst_2023} report a sample of red edge-on disks at $2<z<6$, which are almost invisible in HST ($\lambda_{\rm obs}<1.6\mu$m) but are bright in JWST NIRCam at $\lambda_{\rm obs}=4.4\mu$m due to significant dust attenuation throughout the galaxies. 

Therefore, the half-light radius obtained for galaxies is heavily biased to larger values with respect to the underlying mass distribution due to the dust content and the filter used to observe the galaxies.

\section{Nebular versus Total Dust Attenuation} \label{s:BD_atten}

\begin{figure*}
\begin{tabular}{cc}
    \includegraphics[width=0.7\textwidth]{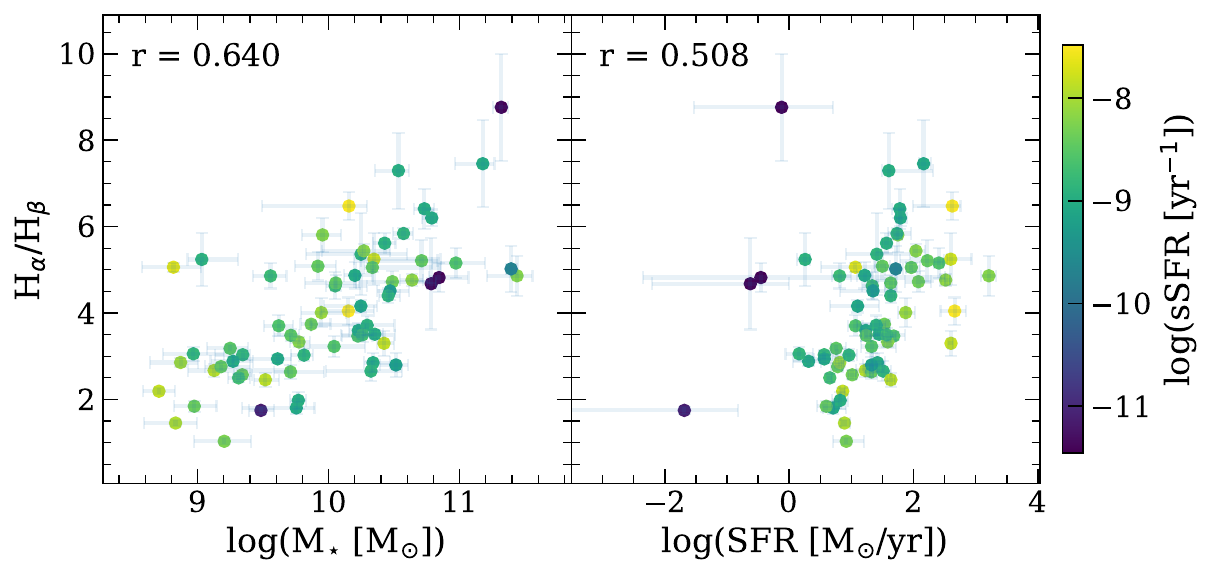} \\
    \includegraphics[width=0.7\textwidth]{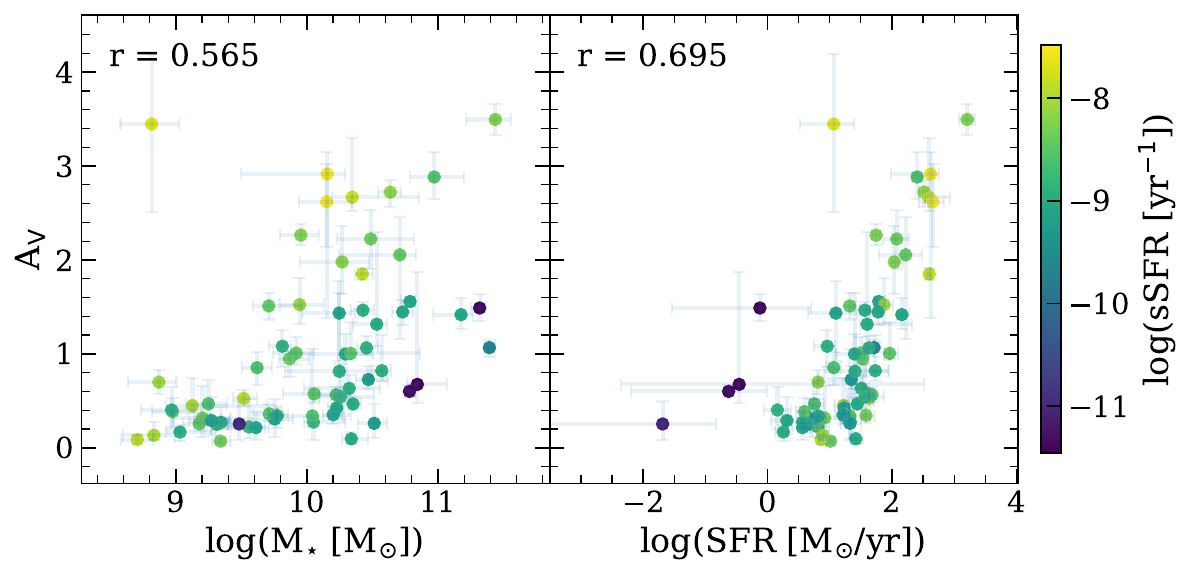} \\ 
\end{tabular}
\caption{The Balmer decrement (H$_{\alpha}$/H$_{\beta}$), tracing the nebular attenuation, is plotted against the stellar mass ($\rm M_{\star}$) and SFR in the top panel, colour-coded by the sSFR. The optical dust attenuation $A_{\rm V}$, probing the attenuation of the stellar continuum from SED fitting, is plotted against the stellar mass and SFR in the bottom panel. The Spearman rank coefficient is written on the figure as 'r', quantifying the strength of the correlation between the X and Y parameters. Both measures of dust attenuation correlate strongly with the stellar mass and SFR; however, the stellar mass correlates more with the Balmer decrement, and the SFR correlates more with $A_{\rm V}$. The dust attenuation versus stellar mass relationship is present when using both the nebular and stellar continuum dust attenuation.} \label{f:BD_corr}
\end{figure*}

\begin{figure}
\centerline{\includegraphics[width=0.5\textwidth]{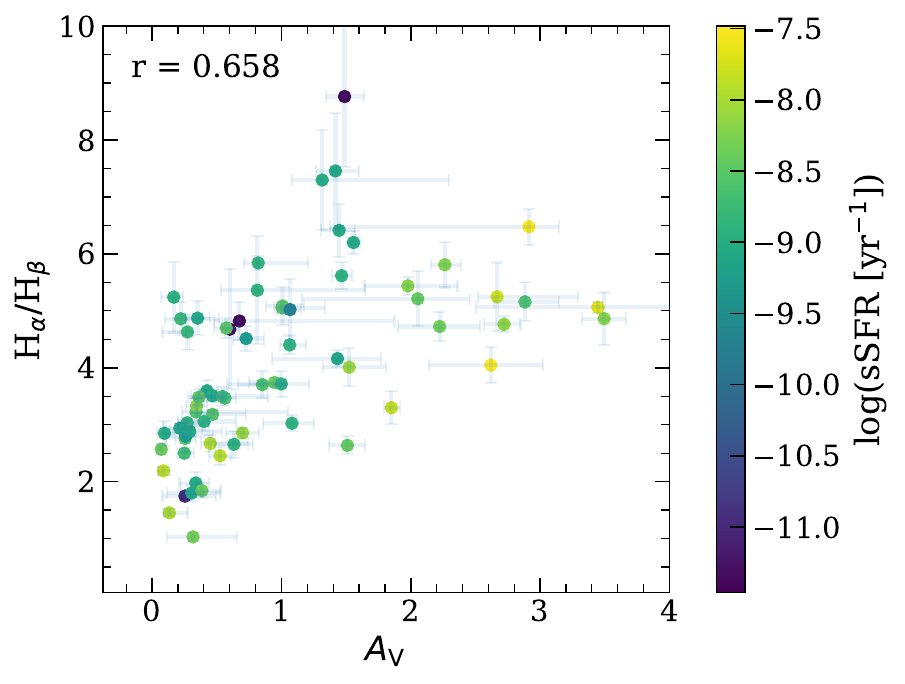}}
\medskip
\caption{The Balmer decrement (H$_{\alpha}$/H$_{\beta}$), measuring the nebular dust attenuation, plotted against the optical dust attenuation (A$_{\rm V}$) from SED fitting, measuring the attenuation of the stellar continuum, colour coded by the sSFR. The Spearman rank coefficient is written on the figure as 'r', quantifying the strength of the correlation between the X and Y parameters. There is a strong correlation between the Balmer decrement and A$_{\rm V}$, indicating that these measures of the dust attenuation agree with each other overall.} \label{f:Av_BD}
\end{figure}

Now, turning our attention to the dust attenuation of emission lines, in this section, we explore how the nebular dust attenuation compares to the stellar continuum dust attenuation. Following the two-component dust model, the emission lines created in nebular regions around young stars are attenuated by the dust in the birth cloud and the dust in the diffuse ISM. The light from young stars is also attenuated by the birth cloud dust and the ISM dust, whereas the light from older stars is only attenuated by the ISM dust. Therefore, the stellar continuum in the optical, composed of both the young and old stellar light, typically undergoes less attenuation than the emission lines. If the stellar continuum is dominated by light from old stars, this discrepancy between the nebular and stellar continuum attenuation will be large. If, instead, the stellar continuum is dominated by young stars, there will be essentially no discrepancy. Previous works have shown that the variation in the dust attenuation of emission lines and of the stellar continuum varies with respect to each other as a function of the SFR \citep{reddy_mosdef_2015}, sSFR \citep{price_direct_2014} and the gas-phase metallicity \citep{shivaei_mosdef_2020}.

In this section, we use the Balmer decrement as a measure of the nebular dust attenuation to compare with the attenuation of the stellar continuum, as measured through $A_{\rm V}$ from the SED fitting with \texttt{Prospector}. We have reliable measurements of the Balmer decrement for 67 galaxies, using signal-to-noise cuts on the Balmer lines (SNR(H$_{\alpha})>25$ and SNR(H$_{\beta})>3$). We first assess the relationship between the Balmer decrement and the stellar population properties, and compare these relationships with those between $A_{\rm V}$ and the stellar population. We then directly compare the Balmer decrement and $A_{\rm V}$. Since the emission lines were marginalised during the SED fitting, the measure of the stellar continuum dust attenuation is independent of the Balmer emission line fluxes.

As in Section~\ref{s:scaling law}, we plot the Balmer decrement against the stellar mass and SFR in the top panel of Figure~\ref{f:BD_corr}. In the bottom panel, we plot the same galaxies now using the optical attenuation ($A_{\rm V}$) from SED fitting. First, looking at the left panel in both plots, we can see that the dust attenuation versus stellar mass relation is present for the nebular dust attenuation, using the Balmer decrement, as well as for the stellar continuum dust attenuation, using $A_{\rm V}$. Due to the high signal-to-noise cuts on the H$_{\alpha}$ emission line, we have mainly highly star-forming galaxies here, and so we are unable to probe the relationship between the dust attenuation and stellar mass for the quenched galaxies.

Considering the right panel of both plots in Figure~\ref{f:BD_corr}, we can see that the SFR strongly correlates with the Balmer decrement and $A_{\rm V}$. Overall, the Spearman rank correlation value "r" shows that the Balmer decrement is slightly more correlated with the stellar mass than the SFR, and $A_{\rm V}$ is slightly more correlated with the SFR than the stellar mass. However, the difference in the strength of these correlations may be driven by random scatter since we are considering a small number of galaxies in this section. A larger sample size would be required to test if this difference in how the Balmer decrement and $A_{\rm V}$ each correlate with the stellar mass and SFR is significant. 

Since the galaxies with high S/N detections of the Balmer lines are mostly star-forming, the light from young stars ($<10$ Myr) will dominate the SED in the rest-frame optical (see Appendix~\ref{App:fl10} for further information). The stellar continuum and the emission lines will then be attenuated by both the birth cloud dust and the ISM dust. Measurements of the Balmer decrement and of $A_{\rm V}$ should therefore be similar, and correlate with the stellar mass and the SFR similarly. We see this result on average in Figure~\ref{f:Av_BD}, where the relationship between the stellar properties and the Balmer decrement is similar to that between the stellar properties and $A_{\rm V}$.

To explore how the Balmer decrement and $A_{\rm V}$ relate to each other, in Figure~\ref{f:Av_BD} we plot the Balmer decrement against A$_{\rm V}$, and we can see that they are strongly correlated, with a Spearman rank coefficient of $0.658$. This correlation indicates that these two probes of dust attenuation agree with each other overall. Following the above argument, most of these galaxies are highly star-forming, and so the nebular dust attenuation will be similar to the dust attenuation of the stellar continuum; therefore, the Balmer decrement and A$_{\rm V}$ from SED fitting should be strongly related.

The values of the Balmer decrement in Figure~\ref{f:BD_corr} and Figure~\ref{f:Av_BD} show that several galaxies have a Balmer decrement less than 2.86, or rather less than the Case B theoretical lower limit. These low Balmer decrements may be caused by small errors in the slit loss corrections or flux calibrations, or they may in fact be true detections of anomolous Balmer emitters, galaxies in which the Case B assumption is invalid. These could be, for example, Lyman-alpha leakers, as proposed by \cite{mcclymont_density-bounded_2024}. The properties of these galaxies will be explored in future work. 

\section{Conclusions}\label{s:conclusion}

In this work, we aim to understand the relationship between the dust attenuation laws and the stellar population parameters of the galaxies and how this may affect the measured morphologies. We obtained photometry and high S/N spectroscopy of a sample of 141 massive ($9<\log(\rm M_{\star}/\rm M_{\odot})<11.5$) galaxies in the COSMOS field using JWST NIRSpec and NIRCam at cosmic noon ($1.7<z<3.5$) from the Blue Jay survey. These observations, along with HST photometry, are fit together using the Bayesian SED fitting code \texttt{Prospector} with a non-parametric SFH and flexible attenuation law. The \cite{charlot_simple_2000} two-component dust model was implemented with the \cite{kriek_dust_2013} attenuation law for the dust in the ISM attenuating the full galaxy (stellar and nebular emission) and a power law for the dust in birth clouds attenuating young ($<10$ Myr) stars and nebular emission. This allowed for a large diversity in the attenuation laws obtained. Morphological fits of deep NIRCam imaging are performed using \texttt{Pysersic} to measure the effective radii and axis ratios of these galaxies in the F150W, F356W and F444W filters. We fit the optical emission lines for these galaxies, providing high S/N measurements of the H$_{\alpha}$ and H$_{\beta}$ emission lines for 67 galaxies.

The high quality of the photometric and spectroscopic data for this sample of galaxies allows for their physical parameters to be well constrained, such as the star-formation history, the effective dust attenuation law and the effective radius across multiple photometric bands. We study the relationship between the attenuation law and the physical and morphological properties of these galaxies, and the main results of this study are as follows:

\begin{enumerate}
    \item The shape and strength of the attenuation law vary with the optical dust attenuation ($A_{\rm V}$), stellar mass and the star formation rate (SFR), presenting large diversity. 
    \item The optical dust attenuation correlates strongly with the stellar mass and the SFR. In addition to this overall trend with stellar mass, we find deviations to the dust attenuation versus stellar mass relationship driven by the star-forming state of the galaxies, since we have measurements for massive quenched galaxies and lower mass starbursts. Massive quiescent galaxies have low dust attenuation compared to their equal-mass, star-forming counterparts, and some lower-mass starbursts have elevated dust attenuation compared to typical star-forming galaxies at a similar stellar mass. For quenched galaxies, the dust in the ISM will be the most significant dust component contributing to the overall dust attenuation law, whereas for starbursts, the birth cloud dust will be the most significant component. Therefore, the stellar mass versus dust attenuation relationship holds for star-forming and quiescent galaxies.
    \item The stellar mass and SFR surface densities correlate more strongly with the dust attenuation than their global counterparts. However, larger sample sizes are required to assess if this is widespread at cosmic noon.
    \item We find no correlation between the axis ratio and the dust attenuation for our galaxies out to a redshift of $z=3.5$. 
    \item The rest-frame optical sizes of our galaxies increase with stellar mass at $1.7<z<3$; however, this flattens off at high stellar masses ($\rm M_{\star}\gtrsim10^{10}~\rm M_{\odot}$) above $z>3$. We measure a flat size-mass relation in the rest-frame NIR across the full redshift range $1.7<z<3.5$.
    \item We find a gradient between the rest-frame optical and NIR sizes. The sizes of the most massive galaxies ($\rm M_{\star}\gtrsim10^{10}~\rm M_{\odot}$) are strongly affected by their dust content, with the galaxies becoming more extended in the rest-optical than the rest-NIR as dust attenuation increases. The elevated dust attenuation in the centre of the galaxies flattens the optical light profile, pushing the effective radius to larger values than in the rest-NIR, where the effect of dust attenuation is less pronounced. On average, dust attenuation increases the size in the rest-optical by $\sim30\%$ compared with the size in the rest-NIR.
    \item For the lower-mass galaxies ($\rm M_{\star}\lesssim10^{10}\rm~M_{\odot}$), we see a large variation in the ratio of rest-frame optical to NIR sizes, likely due to the galaxies undergoing either inside-out growth (more extended in the optical than NIR) or a central starburst (more concentrated in the optical than NIR).
    \item The attenuation of the nebular regions in star-forming galaxies, as traced by the Balmer decrement ($\rm H_{\alpha}/\rm H_{\beta}$), shows similar strong correlations with both the stellar mass and the SFR compared with the attenuation of the stellar continuum measured with \texttt{Prospector}. Due to the high sSFR of the galaxies with clear Balmer decrement measurements, we measure the dust attenuation of the stellar continuum and nebular regions to be strongly related.
\end{enumerate}

Thanks to these recent JWST observations, the complex relationship between dust attenuation, stellar population properties and observed morphologies has become more understandable. We have been able to trace how the stellar population affects the shape of the attenuation law, and the diversity within these attenuation laws demonstrates the need for flexible attenuation laws at this epoch for accurate SED fitting. The strong variations in the observed sizes at different wavelengths show that dust attenuation can drive strong wavelength-dependent size gradients. As more and more objects at cosmic noon are observed with such high S/N from JWST, this study will be extended to larger sample sizes to cement our findings on a population level. 

\section{Acknowledgements}

GM acknowledges support from the Science and Technology Facilities Council (STFC), EJN acknowledges support from JWST-GO-01810 and JWST-GO-04106, and SB, LB, and AHK acknowledge support from ERC grant 101076080 "Red Cardinal".



\bibliographystyle{mnras}
\bibliography{references_17042025} 

\begin{thebibliography}{}
\makeatletter
\relax
\def\mn@urlcharsother{\let\do\@makeother \do\$\do\&\do\#\do\^\do\_\do\%\do\~}
\def\mn@doi{\begingroup\mn@urlcharsother \@ifnextchar [ {\mn@doi@} {\mn@doi@[]}}
\def\mn@doi@[#1]#2{\def\@tempa{#1}\ifx\@tempa\@empty \href {http://dx.doi.org/#2} {doi:#2}\else \href {http://dx.doi.org/#2} {#1}\fi \endgroup}
\def\mn@eprint#1#2{\mn@eprint@#1:#2::\@nil}
\def\mn@eprint@arXiv#1{\href {http://arxiv.org/abs/#1} {{\tt arXiv:#1}}}
\def\mn@eprint@dblp#1{\href {http://dblp.uni-trier.de/rec/bibtex/#1.xml} {dblp:#1}}
\def\mn@eprint@#1:#2:#3:#4\@nil{\def\@tempa {#1}\def\@tempb {#2}\def\@tempc {#3}\ifx \@tempc \@empty \let \@tempc \@tempb \let \@tempb \@tempa \fi \ifx \@tempb \@empty \def\@tempb {arXiv}\fi \@ifundefined {mn@eprint@\@tempb}{\@tempb:\@tempc}{\expandafter \expandafter \csname mn@eprint@\@tempb\endcsname \expandafter{\@tempc}}}

\bibitem[\protect\citeauthoryear{Allen et~al.,}{Allen et~al.}{2024}]{allen_galaxy_2024}
Allen N.,  et~al., 2024, Galaxy {Size} and {Mass} {Build}-up in the {First} 2 {Gyrs} of {Cosmic} {History} from {Multi}-{Wavelength} {JWST} {NIRCam} {Imaging}, \mn@doi{10.48550/arXiv.2410.16354}, \url {https://ui.adsabs.harvard.edu/abs/2024arXiv241016354A}

\bibitem[\protect\citeauthoryear{Alsing, Thorp, Deger, Peiris, Leistedt, Mortlock  \& Leja}{Alsing et~al.}{2024}]{alsing_pop-cosmos_2024}
Alsing J.,  Thorp S.,  Deger S.,  Peiris H.~V.,  Leistedt B.,  Mortlock D.,   Leja J.,  2024, \mn@doi [The Astrophysical Journal Supplement Series] {10.3847/1538-4365/ad5c69}, 274, 12

\bibitem[\protect\citeauthoryear{Baker, Maiolino, Bluck, Lin, Ellison, Belfiore, Pan  \& Thorp}{Baker et~al.}{2022}]{baker_almaquest_2022}
Baker W.~M.,  Maiolino R.,  Bluck A. F.~L.,  Lin L.,  Ellison S.~L.,  Belfiore F.,  Pan H.-A.,   Thorp M.,  2022, \mn@doi [Monthly Notices of the Royal Astronomical Society] {10.1093/mnras/stab3672}, 510, 3622

\bibitem[\protect\citeauthoryear{Belli et~al.,}{Belli et~al.}{2023}]{belli_massive_2023}
Belli S.,  et~al., 2023, Massive and {Multiphase} {Gas} {Outflow} in a {Quenching} {Galaxy} at z=2.445, \mn@doi{10.48550/arXiv.2308.05795}, \url {https://ui.adsabs.harvard.edu/abs/2023arXiv230805795B}

\bibitem[\protect\citeauthoryear{Belli et~al.,}{Belli et~al.}{2024}]{belli_star_2024}
Belli S.,  et~al., 2024, \mn@doi [Nature] {10.1038/s41586-024-07412-1}, 630, 54

\bibitem[\protect\citeauthoryear{Blitz \& Shu}{Blitz \& Shu}{1980}]{blitz_origin_1980}
Blitz L.,  Shu F.~H.,  1980, \mn@doi [The Astrophysical Journal] {10.1086/157968}, 238, 148

\bibitem[\protect\citeauthoryear{Bluck, Maiolino, Sánchez, Ellison, Thorp, Piotrowska, Teimoorinia  \& Bundy}{Bluck et~al.}{2020}]{bluck_are_2020}
Bluck A. F.~L.,  Maiolino R.,  Sánchez S.~F.,  Ellison S.~L.,  Thorp M.~D.,  Piotrowska J.~M.,  Teimoorinia H.,   Bundy K.~A.,  2020, \mn@doi [Monthly Notices of the Royal Astronomical Society] {10.1093/mnras/stz3264}, 492, 96

\bibitem[\protect\citeauthoryear{Bradley et~al.,}{Bradley et~al.}{2024}]{bradley_astropyphotutils_2024}
Bradley L.,  et~al., 2024, astropy/photutils: 2.0.2, \mn@doi{10.5281/zenodo.13989456}, \url {https://zenodo.org/records/13989456}

\bibitem[\protect\citeauthoryear{Brinchmann, Charlot, White, Tremonti, Kauffmann, Heckman  \& Brinkmann}{Brinchmann et~al.}{2004}]{brinchmann_physical_2004}
Brinchmann J.,  Charlot S.,  White S. D.~M.,  Tremonti C.,  Kauffmann G.,  Heckman T.,   Brinkmann J.,  2004, \mn@doi [Monthly Notices of the Royal Astronomical Society] {10.1111/j.1365-2966.2004.07881.x}, 351, 1151

\bibitem[\protect\citeauthoryear{Bugiani et~al.,}{Bugiani et~al.}{2024}]{bugiani_agn_2024}
Bugiani L.,  et~al., 2024, {AGN} {Feedback} in {Quiescent} {Galaxies} at {Cosmic} {Noon} {Traced} by {Ionized} {Gas} {Emission}, \url {http://arxiv.org/abs/2406.08547}

\bibitem[\protect\citeauthoryear{Calzetti, Kinney  \& Storchi-Bergmann}{Calzetti et~al.}{1994}]{calzetti_dust_1994}
Calzetti D.,  Kinney A.~L.,   Storchi-Bergmann T.,  1994, \mn@doi [The Astrophysical Journal] {10.1086/174346}, 429, 582

\bibitem[\protect\citeauthoryear{Calzetti, Armus, Bohlin, Kinney, Koornneef  \& Storchi-Bergmann}{Calzetti et~al.}{2000}]{calzetti_dust_2000}
Calzetti D.,  Armus L.,  Bohlin R.~C.,  Kinney A.~L.,  Koornneef J.,   Storchi-Bergmann T.,  2000, \mn@doi [The Astrophysical Journal] {10.1086/308692}, 533, 682

\bibitem[\protect\citeauthoryear{Cardelli, Clayton  \& Mathis}{Cardelli et~al.}{1989}]{cardelli_relationship_1989}
Cardelli J.~A.,  Clayton G.~C.,   Mathis J.~S.,  1989, \mn@doi [The Astrophysical Journal] {10.1086/167900}, 345, 245

\bibitem[\protect\citeauthoryear{Cargile, Conroy, Johnson, Ting, Bonaca, Dotter  \& Speagle}{Cargile et~al.}{2020}]{cargile_minesweeper_2020}
Cargile P.~A.,  Conroy C.,  Johnson B.~D.,  Ting Y.-S.,  Bonaca A.,  Dotter A.,   Speagle J.~S.,  2020, \mn@doi [The Astrophysical Journal] {10.3847/1538-4357/aba43b}, 900, 28

\bibitem[\protect\citeauthoryear{Cebrián \& Trujillo}{Cebrián \& Trujillo}{2014}]{cebrian_effect_2014}
Cebrián M.,  Trujillo I.,  2014, \mn@doi [Monthly Notices of the Royal Astronomical Society] {10.1093/mnras/stu1375}, 444, 682

\bibitem[\protect\citeauthoryear{Chabrier}{Chabrier}{2003}]{chabrier_galactic_2003}
Chabrier G.,  2003, \mn@doi [Publications of the Astronomical Society of the Pacific] {10.1086/376392}, 115, 763

\bibitem[\protect\citeauthoryear{Chakraborty, Ferland, Chatzikos, Guzmán  \& Su}{Chakraborty et~al.}{2021}]{chakraborty_x-ray_2021}
Chakraborty P.,  Ferland G.~J.,  Chatzikos M.,  Guzmán F.,   Su Y.,  2021, \mn@doi [The Astrophysical Journal] {10.3847/1538-4357/abed4a}, 912, 26

\bibitem[\protect\citeauthoryear{Charlot \& Fall}{Charlot \& Fall}{2000}]{charlot_simple_2000}
Charlot S.,  Fall S.~M.,  2000, \mn@doi [The Astrophysical Journal] {10.1086/309250}, 539, 718

\bibitem[\protect\citeauthoryear{Chevallard, Charlot, Wandelt  \& Wild}{Chevallard et~al.}{2013}]{chevallard_insights_2013}
Chevallard J.,  Charlot S.,  Wandelt B.,   Wild V.,  2013, \mn@doi [Monthly Notices of the Royal Astronomical Society] {10.1093/mnras/stt523}, 432, 2061

\bibitem[\protect\citeauthoryear{Chevance et~al.,}{Chevance et~al.}{2020}]{chevance_lifecycle_2020}
Chevance M.,  et~al., 2020, \mn@doi [Monthly Notices of the Royal Astronomical Society] {10.1093/mnras/stz3525}, 493, 2872

\bibitem[\protect\citeauthoryear{Choi, Dotter, Conroy, Cantiello, Paxton  \& Johnson}{Choi et~al.}{2016}]{choi_mesa_2016}
Choi J.,  Dotter A.,  Conroy C.,  Cantiello M.,  Paxton B.,   Johnson B.~D.,  2016, \mn@doi [The Astrophysical Journal] {10.3847/0004-637X/823/2/102}, 823, 102

\bibitem[\protect\citeauthoryear{Conroy \& Gunn}{Conroy \& Gunn}{2010}]{conroy_propagation_2010}
Conroy C.,  Gunn J.~E.,  2010, \mn@doi [The Astrophysical Journal] {10.1088/0004-637X/712/2/833}, 712, 833

\bibitem[\protect\citeauthoryear{Conroy, Gunn  \& White}{Conroy et~al.}{2009}]{conroy_propagation_2009}
Conroy C.,  Gunn J.~E.,   White M.,  2009, \mn@doi [The Astrophysical Journal] {10.1088/0004-637X/699/1/486}, 699, 486

\bibitem[\protect\citeauthoryear{Davies et~al.,}{Davies et~al.}{2024}]{davies_jwst_2024}
Davies R.~L.,  et~al., 2024, \mn@doi [Monthly Notices of the Royal Astronomical Society] {10.1093/mnras/stae327}, 528, 4976

\bibitem[\protect\citeauthoryear{Draine}{Draine}{2003}]{draine_interstellar_2003}
Draine B.~T.,  2003, \mn@doi [Annual Review of Astronomy and Astrophysics] {10.1146/annurev.astro.41.011802.094840}, 41, 241

\bibitem[\protect\citeauthoryear{Draine}{Draine}{2011}]{draine_physics_2011}
Draine B.~T.,  2011, Physics of the {Interstellar} and {Intergalactic} {Medium}.
\url {https://ui.adsabs.harvard.edu/abs/2011piim.book.....D}

\bibitem[\protect\citeauthoryear{Dubois, Peirani, Pichon, Devriendt, Gavazzi, Welker  \& Volonteri}{Dubois et~al.}{2016}]{dubois_horizon-agn_2016}
Dubois Y.,  Peirani S.,  Pichon C.,  Devriendt J.,  Gavazzi R.,  Welker C.,   Volonteri M.,  2016, \mn@doi [Monthly Notices of the Royal Astronomical Society] {10.1093/mnras/stw2265}, 463, 3948

\bibitem[\protect\citeauthoryear{Fanelli, O'Connell  \& Thuan}{Fanelli et~al.}{1988}]{fanelli_spectral_1988}
Fanelli M.~N.,  O'Connell R.~W.,   Thuan T.~X.,  1988, \mn@doi [The Astrophysical Journal] {10.1086/166869}, 334, 665

\bibitem[\protect\citeauthoryear{Ferruit et~al.,}{Ferruit et~al.}{2022}]{ferruit_near-infrared_2022}
Ferruit P.,  et~al., 2022, \mn@doi [Astronomy \& Astrophysics] {10.1051/0004-6361/202142673}, 661, A81

\bibitem[\protect\citeauthoryear{Foreman-Mackey, Hogg, Lang  \& Goodman}{Foreman-Mackey et~al.}{2013}]{foreman-mackey_emcee_2013}
Foreman-Mackey D.,  Hogg D.~W.,  Lang D.,   Goodman J.,  2013, \mn@doi [Publications of the Astronomical Society of the Pacific] {10.1086/670067}, 125, 306

\bibitem[\protect\citeauthoryear{Frankel, Sanders, Rix, Ting  \& Ness}{Frankel et~al.}{2019}]{frankel_inside-out_2019}
Frankel N.,  Sanders J.,  Rix H.-W.,  Ting Y.-S.,   Ness M.,  2019, \mn@doi [The Astrophysical Journal] {10.3847/1538-4357/ab4254}, 884, 99

\bibitem[\protect\citeauthoryear{Förster~Schreiber et~al.,}{Förster~Schreiber et~al.}{2011}]{forster_schreiber_constraints_2011}
Förster~Schreiber N.~M.,  et~al., 2011, \mn@doi [The Astrophysical Journal] {10.1088/0004-637X/739/1/45}, 739, 45

\bibitem[\protect\citeauthoryear{Gall, Hjorth  \& Andersen}{Gall et~al.}{2011}]{gall_production_2011}
Gall C.,  Hjorth J.,   Andersen A.~C.,  2011, \mn@doi [The Astronomy and Astrophysics Review] {10.1007/s00159-011-0043-7}, 19, 43

\bibitem[\protect\citeauthoryear{Gallerani et~al.,}{Gallerani et~al.}{2010}]{gallerani_extinction_2010}
Gallerani S.,  et~al., 2010, \mn@doi [Astronomy and Astrophysics] {10.1051/0004-6361/201014721}, 523, A85

\bibitem[\protect\citeauthoryear{Garn \& Best}{Garn \& Best}{2010}]{garn_predicting_2010}
Garn T.,  Best P.~N.,  2010, \mn@doi [Monthly Notices of the Royal Astronomical Society] {10.1111/j.1365-2966.2010.17321.x}, 409, 421

\bibitem[\protect\citeauthoryear{Gillett, Jones, Merrill  \& Stein}{Gillett et~al.}{1975}]{gillett_anisotropy_1975}
Gillett F.~C.,  Jones T.~W.,  Merrill K.~M.,   Stein W.~A.,  1975, Astronomy and Astrophysics, 45, 77

\bibitem[\protect\citeauthoryear{Gordon, Calzetti  \& Witt}{Gordon et~al.}{1997}]{gordon_dust_1997}
Gordon K.~D.,  Calzetti D.,   Witt A.~N.,  1997, \mn@doi [The Astrophysical Journal] {10.1086/304654}, 487, 625

\bibitem[\protect\citeauthoryear{Gordon, Clayton, Misselt, Landolt  \& Wolff}{Gordon et~al.}{2003}]{gordon_quantitative_2003}
Gordon K.~D.,  Clayton G.~C.,  Misselt K.~A.,  Landolt A.~U.,   Wolff M.~J.,  2003, \mn@doi [The Astrophysical Journal] {10.1086/376774}, 594, 279

\bibitem[\protect\citeauthoryear{Hahn \& Melchior}{Hahn \& Melchior}{2024}]{hahn_inhomogeneous_2024}
Hahn C.,  Melchior P.,  2024, Inhomogeneous {Dust} {Biases} {Photometric} {Redshifts} and {Stellar} {Masses} for {LSST}, \mn@doi{10.48550/arXiv.2409.19054}, \url {https://ui.adsabs.harvard.edu/abs/2024arXiv240919054H}

\bibitem[\protect\citeauthoryear{Hayden-Pawson et~al.,}{Hayden-Pawson et~al.}{2022}]{hayden-pawson_klever_2022}
Hayden-Pawson C.,  et~al., 2022, \mn@doi [Monthly Notices of the Royal Astronomical Society] {10.1093/mnras/stac584}, 512, 2867

\bibitem[\protect\citeauthoryear{Hou, Hirashita, Nagamine, Aoyama  \& Shimizu}{Hou et~al.}{2017}]{hou_evolution_2017}
Hou K.-C.,  Hirashita H.,  Nagamine K.,  Aoyama S.,   Shimizu I.,  2017, \mn@doi [Monthly Notices of the Royal Astronomical Society] {10.1093/mnras/stx877}, 469, 870

\bibitem[\protect\citeauthoryear{Inami et~al.,}{Inami et~al.}{2022}]{inami_alma_2022}
Inami H.,  et~al., 2022, \mn@doi [Monthly Notices of the Royal Astronomical Society] {10.1093/mnras/stac1779}, 515, 3126

\bibitem[\protect\citeauthoryear{Jackson, Kaviraj, Martin, Devriendt, Noakes-Kettel, Silk, Ogle  \& Dubois}{Jackson et~al.}{2022}]{jackson_extremely_2022}
Jackson R.~A.,  Kaviraj S.,  Martin G.,  Devriendt J. E.~G.,  Noakes-Kettel E.~A.,  Silk J.,  Ogle P.,   Dubois Y.,  2022, \mn@doi [Monthly Notices of the Royal Astronomical Society] {10.1093/mnras/stac058}, 511, 607

\bibitem[\protect\citeauthoryear{Ji et~al.,}{Ji et~al.}{2023}]{ji_jades_2023}
Ji Z.,  et~al., 2023, {JADES} + {JEMS}: {A} {Detailed} {Look} at the {Buildup} of {Central} {Stellar} {Cores} and {Suppression} of {Star} {Formation} in {Galaxies} at {Redshifts} 3 {\textless} z {\textless} 4.5, \mn@doi{10.48550/arXiv.2305.18518}, \url {http://arxiv.org/abs/2305.18518}

\bibitem[\protect\citeauthoryear{Johnson, Leja, Conroy  \& Speagle}{Johnson et~al.}{2021}]{johnson_stellar_2021}
Johnson B.~D.,  Leja J.,  Conroy C.,   Speagle J.~S.,  2021, \mn@doi [The Astrophysical Journal Supplement Series] {10.3847/1538-4365/abef67}, 254, 22

\bibitem[\protect\citeauthoryear{Kawinwanichakij et~al.,}{Kawinwanichakij et~al.}{2021}]{kawinwanichakij_hyper_2021}
Kawinwanichakij L.,  et~al., 2021, \mn@doi [The Astrophysical Journal] {10.3847/1538-4357/ac1f21}, 921, 38

\bibitem[\protect\citeauthoryear{Kemper, Vriend  \& Tielens}{Kemper et~al.}{2004}]{kemper_absence_2004}
Kemper F.,  Vriend W.~J.,   Tielens A. G. G.~M.,  2004, \mn@doi [The Astrophysical Journal] {10.1086/421339}, 609, 826

\bibitem[\protect\citeauthoryear{Kriek \& Conroy}{Kriek \& Conroy}{2013}]{kriek_dust_2013}
Kriek M.,  Conroy C.,  2013, \mn@doi [The Astrophysical Journal] {10.1088/2041-8205/775/1/L16}, 775, L16

\bibitem[\protect\citeauthoryear{Laporte et~al.,}{Laporte et~al.}{2017}]{laporte_dust_2017}
Laporte N.,  et~al., 2017, \mn@doi [The Astrophysical Journal Letters] {10.3847/2041-8213/aa62aa}, 837, L21

\bibitem[\protect\citeauthoryear{Leja, Carnall, Johnson, Conroy  \& Speagle}{Leja et~al.}{2019a}]{leja_how_2019}
Leja J.,  Carnall A.~C.,  Johnson B.~D.,  Conroy C.,   Speagle J.~S.,  2019a, \mn@doi [The Astrophysical Journal] {10.3847/1538-4357/ab133c}, 876, 3

\bibitem[\protect\citeauthoryear{Leja et~al.,}{Leja et~al.}{2019b}]{leja_older_2019}
Leja J.,  et~al., 2019b, \mn@doi [The Astrophysical Journal] {10.3847/1538-4357/ab1d5a}, 877, 140

\bibitem[\protect\citeauthoryear{Leja et~al.,}{Leja et~al.}{2022}]{leja_new_2022}
Leja J.,  et~al., 2022, \mn@doi [The Astrophysical Journal] {10.3847/1538-4357/ac887d}, 936, 165

\bibitem[\protect\citeauthoryear{Lian, Yan, Blanton  \& Kong}{Lian et~al.}{2017}]{lian_inside-out_2017}
Lian J.,  Yan R.,  Blanton M.,   Kong X.,  2017, \mn@doi [Monthly Notices of the Royal Astronomical Society] {10.1093/mnras/stx2216}, 472, 4679

\bibitem[\protect\citeauthoryear{Lilly \& Carollo}{Lilly \& Carollo}{2016}]{lilly_surface_2016}
Lilly S.~J.,  Carollo C.~M.,  2016, \mn@doi [The Astrophysical Journal] {10.3847/0004-637X/833/1/1}, 833, 1

\bibitem[\protect\citeauthoryear{Lin et~al.,}{Lin et~al.}{2019}]{lin_almaquest_2019}
Lin L.,  et~al., 2019, \mn@doi [The Astrophysical Journal] {10.3847/2041-8213/ab4815}, 884, L33

\bibitem[\protect\citeauthoryear{Lin, Yang, Li  \& Witstok}{Lin et~al.}{2025}]{lin_polycyclic_2025}
Lin Q.,  Yang X.,  Li A.,   Witstok J.,  2025, \mn@doi [Astronomy and Astrophysics] {10.1051/0004-6361/202452372}, 694, A84

\bibitem[\protect\citeauthoryear{Looser, D'Eugenio, Piotrowska, Belfiore, Maiolino, Cappellari, Baker  \& Tacchella}{Looser et~al.}{2024}]{looser_stellar_2024}
Looser T.~J.,  D'Eugenio F.,  Piotrowska J.~M.,  Belfiore F.,  Maiolino R.,  Cappellari M.,  Baker W.~M.,   Tacchella S.,  2024, The stellar {Fundamental} {Metallicity} {Relation}: the correlation between stellar mass, star-formation rate and stellar metallicity, \url {http://arxiv.org/abs/2401.08769}

\bibitem[\protect\citeauthoryear{Lorenz et~al.,}{Lorenz et~al.}{2023}]{lorenz_updated_2023}
Lorenz B.,  et~al., 2023, \mn@doi [The Astrophysical Journal] {10.3847/1538-4357/accdd1}, 951, 29

\bibitem[\protect\citeauthoryear{Lyu et~al.,}{Lyu et~al.}{2025}]{lyu_primer_2025}
Lyu Y.,  et~al., 2025, \mn@doi [Astronomy and Astrophysics] {10.1051/0004-6361/202451067}, 693, A313

\bibitem[\protect\citeauthoryear{Maheson, Maiolino, Curti, Sanders, Tacchella  \& Sandles}{Maheson et~al.}{2024}]{maheson_unravelling_2024}
Maheson G.,  Maiolino R.,  Curti M.,  Sanders R.,  Tacchella S.,   Sandles L.,  2024, \mn@doi [Monthly Notices of the Royal Astronomical Society] {10.1093/mnras/stad3685}, 527, 8213

\bibitem[\protect\citeauthoryear{Mannucci, Cresci, Maiolino, Marconi  \& Gnerucci}{Mannucci et~al.}{2010}]{mannucci_fundamental_2010}
Mannucci F.,  Cresci G.,  Maiolino R.,  Marconi A.,   Gnerucci A.,  2010, \mn@doi [Monthly Notices of the Royal Astronomical Society] {10.1111/j.1365-2966.2010.17291.x}, 408, 2115

\bibitem[\protect\citeauthoryear{Markov, Gallerani, Pallottini, Sommovigo, Carniani, Ferrara, Parlanti  \& Mascia}{Markov et~al.}{2023}]{markov_dust_2023}
Markov V.,  Gallerani S.,  Pallottini A.,  Sommovigo L.,  Carniani S.,  Ferrara A.,  Parlanti E.,   Mascia F.~D.,  2023, \mn@doi [Astronomy \& Astrophysics] {10.1051/0004-6361/202346723}, 679, A12

\bibitem[\protect\citeauthoryear{Markov, Gallerani, Ferrara, Pallottini, Parlanti, Di~Mascia, Sommovigo  \& Kohandel}{Markov et~al.}{2024}]{markov_evolution_2024}
Markov V.,  Gallerani S.,  Ferrara A.,  Pallottini A.,  Parlanti E.,  Di~Mascia F.,  Sommovigo L.,   Kohandel M.,  2024, \mn@doi [Nature Astronomy] {10.1038/s41550-024-02426-1}

\bibitem[\protect\citeauthoryear{Martorano, van~der Wel, Baes, Bell, Brammer, Franx  \& Nersesian}{Martorano et~al.}{2024}]{martorano_sizemass_2024}
Martorano M.,  van~der Wel A.,  Baes M.,  Bell E.~F.,  Brammer G.,  Franx M.,   Nersesian A.,  2024, \mn@doi [The Astrophysical Journal] {10.3847/1538-4357/ad5c6a}, 972, 134

\bibitem[\protect\citeauthoryear{Mascia, Gallerani, Ferrara, Pallottini, Maiolino, Carniani  \& D'Odorico}{Mascia et~al.}{2021}]{mascia_dust_2021}
Mascia F.~D.,  Gallerani S.,  Ferrara A.,  Pallottini A.,  Maiolino R.,  Carniani S.,   D'Odorico V.,  2021, \mn@doi [Monthly Notices of the Royal Astronomical Society] {10.1093/mnras/stab1876}, 506, 3946

\bibitem[\protect\citeauthoryear{Matharu et~al.,}{Matharu et~al.}{2023}]{matharu_first_2023}
Matharu J.,  et~al., 2023, \mn@doi [The Astrophysical Journal] {10.3847/2041-8213/acd1db}, 949, L11

\bibitem[\protect\citeauthoryear{McClymont et~al.,}{McClymont et~al.}{2024}]{mcclymont_density-bounded_2024}
McClymont W.,  et~al., 2024, The density-bounded twilight of starbursts in the early {Universe}, \mn@doi{10.48550/arXiv.2405.15859}, \url {http://arxiv.org/abs/2405.15859}

\bibitem[\protect\citeauthoryear{McClymont et~al.,}{McClymont et~al.}{2025a}]{mcclymont_thesan-zoom_2025}
McClymont W.,  et~al., 2025a, The {THESAN}-{ZOOM} project: {Burst}, quench, repeat -- unveiling the evolution of high-redshift galaxies along the star-forming main sequence, \mn@doi{10.48550/arXiv.2503.00106}, \url {https://ui.adsabs.harvard.edu/abs/2025arXiv250300106M}

\bibitem[\protect\citeauthoryear{McClymont et~al.,}{McClymont et~al.}{2025b}]{mcclymont_thesan-zoom_2025-1}
McClymont W.,  et~al., 2025b, The {THESAN}-{ZOOM} project: central starbursts and inside-out quenching govern galaxy sizes in the early {Universe}, \mn@doi{10.48550/arXiv.2503.04894}, \url {http://arxiv.org/abs/2503.04894}

\bibitem[\protect\citeauthoryear{Micelotta, Matsuura  \& Sarangi}{Micelotta et~al.}{2018}]{micelotta_dust_2018}
Micelotta E.~R.,  Matsuura M.,   Sarangi A.,  2018, \mn@doi [Space Science Reviews] {10.1007/s11214-018-0484-7}, 214, 53

\bibitem[\protect\citeauthoryear{Miller et~al.,}{Miller et~al.}{2024}]{miller_jwst_2024}
Miller T.~B.,  et~al., 2024, {JWST} {UNCOVERs} the {Optical} {Size} - {Stellar} {Mass} {Relation} at \$4, \mn@doi{10.48550/arXiv.2412.06957}, \url {https://ui.adsabs.harvard.edu/abs/2024arXiv241206957M}

\bibitem[\protect\citeauthoryear{Momcheva et~al.,}{Momcheva et~al.}{2016}]{momcheva_3d-hst_2016}
Momcheva I.~G.,  et~al., 2016, \mn@doi [The Astrophysical Journal Supplement Series] {10.3847/0067-0049/225/2/27}, 225, 27

\bibitem[\protect\citeauthoryear{Mosleh, Hosseinnejad, Hosseini-ShahiSavandi  \& Tacchella}{Mosleh et~al.}{2020}]{mosleh_galaxy_2020}
Mosleh M.,  Hosseinnejad S.,  Hosseini-ShahiSavandi S.~Z.,   Tacchella S.,  2020, \mn@doi [The Astrophysical Journal] {10.3847/1538-4357/abc7cc}, 905, 170

\bibitem[\protect\citeauthoryear{Mowla et~al.,}{Mowla et~al.}{2019}]{mowla_cosmos-dash_2019}
Mowla L.~A.,  et~al., 2019, \mn@doi [The Astrophysical Journal] {10.3847/1538-4357/ab290a}, 880, 57

\bibitem[\protect\citeauthoryear{Muñoz-Mateos, Gil~de Paz, Boissier, Zamorano, Jarrett, Gallego  \& Madore}{Muñoz-Mateos et~al.}{2007}]{munoz-mateos_specific_2007}
Muñoz-Mateos J.~C.,  Gil~de Paz A.,  Boissier S.,  Zamorano J.,  Jarrett T.,  Gallego J.,   Madore B.~F.,  2007, \mn@doi [The Astrophysical Journal] {10.1086/511812}, 658, 1006

\bibitem[\protect\citeauthoryear{Naab, Johansson  \& Ostriker}{Naab et~al.}{2009}]{naab_minor_2009}
Naab T.,  Johansson P.~H.,   Ostriker J.~P.,  2009, \mn@doi [The Astrophysical Journal] {10.1088/0004-637X/699/2/L178}, 699, L178

\bibitem[\protect\citeauthoryear{Nagaraj, Forbes, Leja, Foreman-Mackey  \& Hayward}{Nagaraj et~al.}{2022}]{nagaraj_bayesian_2022}
Nagaraj G.,  Forbes J.~C.,  Leja J.,  Foreman-Mackey D.,   Hayward C.~C.,  2022, \mn@doi [The Astrophysical Journal] {10.3847/1538-4357/ac6c80}, 932, 54

\bibitem[\protect\citeauthoryear{Narayanan, Conroy, Davé, Johnson  \& Popping}{Narayanan et~al.}{2018}]{narayanan_theory_2018}
Narayanan D.,  Conroy C.,  Davé R.,  Johnson B.~D.,   Popping G.,  2018, \mn@doi [The Astrophysical Journal] {10.3847/1538-4357/aaed25}, 869, 70

\bibitem[\protect\citeauthoryear{Nedkova et~al.,}{Nedkova et~al.}{2024}]{nedkova_uvcandels_2024}
Nedkova K.~V.,  et~al., 2024, \mn@doi [The Astrophysical Journal] {10.3847/1538-4357/ad4ede}, 970, 188

\bibitem[\protect\citeauthoryear{Nelson et~al.,}{Nelson et~al.}{2016a}]{nelson_spatially_2016}
Nelson E.~J.,  et~al., 2016a, \mn@doi [The Astrophysical Journal] {10.3847/2041-8205/817/1/L9}, 817, L9

\bibitem[\protect\citeauthoryear{Nelson et~al.,}{Nelson et~al.}{2016b}]{nelson_where_2016}
Nelson E.~J.,  et~al., 2016b, \mn@doi [The Astrophysical Journal] {10.3847/0004-637X/828/1/27}, 828, 27

\bibitem[\protect\citeauthoryear{Nelson et~al.,}{Nelson et~al.}{2021}]{nelson_spatially_2021}
Nelson E.~J.,  et~al., 2021, \mn@doi [Monthly Notices of the Royal Astronomical Society] {10.1093/mnras/stab2131}, 508, 219

\bibitem[\protect\citeauthoryear{Nelson et~al.,}{Nelson et~al.}{2023}]{nelson_jwst_2023}
Nelson E.~J.,  et~al., 2023, \mn@doi [The Astrophysical Journal] {10.3847/2041-8213/acc1e1}, 948, L18

\bibitem[\protect\citeauthoryear{Noll, Burgarella, Giovannoli, Buat, Marcillac  \& Muñoz-Mateos}{Noll et~al.}{2009}]{noll_analysis_2009}
Noll S.,  Burgarella D.,  Giovannoli E.,  Buat V.,  Marcillac D.,   Muñoz-Mateos J.~C.,  2009, \mn@doi [Astronomy and Astrophysics] {10.1051/0004-6361/200912497}, 507, 1793

\bibitem[\protect\citeauthoryear{Orellana et~al.,}{Orellana et~al.}{2017}]{orellana_molecular_2017}
Orellana G.,  et~al., 2017, \mn@doi [Astronomy \& Astrophysics] {10.1051/0004-6361/201629009}, 602, A68

\bibitem[\protect\citeauthoryear{Padilla \& Strauss}{Padilla \& Strauss}{2008}]{padilla_shapes_2008}
Padilla N.~D.,  Strauss M.~A.,  2008, \mn@doi [Monthly Notices of the Royal Astronomical Society] {10.1111/j.1365-2966.2008.13480.x}, 388, 1321

\bibitem[\protect\citeauthoryear{Papoular \& Papoular}{Papoular \& Papoular}{2009}]{papoular_polycrystalline_2009}
Papoular R.~J.,  Papoular R.,  2009, \mn@doi [Monthly Notices of the Royal Astronomical Society] {10.1111/j.1365-2966.2009.14484.x}, 394, 2175

\bibitem[\protect\citeauthoryear{Papovich, Dickinson  \& Ferguson}{Papovich et~al.}{2001}]{papovich_stellar_2001}
Papovich C.,  Dickinson M.,   Ferguson H.~C.,  2001, \mn@doi [The Astrophysical Journal] {10.1086/322412}, 559, 620

\bibitem[\protect\citeauthoryear{Park et~al.,}{Park et~al.}{2024}]{park_widespread_2024}
Park M.,  et~al., 2024, Widespread rapid quenching at cosmic noon revealed by {JWST} deep spectroscopy, \mn@doi{10.48550/arXiv.2404.17945}, \url {https://ui.adsabs.harvard.edu/abs/2024arXiv240417945P}

\bibitem[\protect\citeauthoryear{Pasha \& Miller}{Pasha \& Miller}{2023}]{pasha_pysersic_2023}
Pasha I.,  Miller T.~B.,  2023, pysersic: {A} {Python} package for determining galaxy structural properties via {Bayesian} inference, accelerated with jax, \mn@doi{10.48550/arXiv.2306.05454}, \url {http://arxiv.org/abs/2306.05454}

\bibitem[\protect\citeauthoryear{Perrin, Sivaramakrishnan, Lajoie, Elliott, Pueyo, Ravindranath  \& Albert}{Perrin et~al.}{2014}]{perrin_updated_2014}
Perrin M.~D.,  Sivaramakrishnan A.,  Lajoie C.-P.,  Elliott E.,  Pueyo L.,  Ravindranath S.,   Albert L.,  2014. p. 91433X, \mn@doi{10.1117/12.2056689}, \url {https://ui.adsabs.harvard.edu/abs/2014SPIE.9143E..3XP}

\bibitem[\protect\citeauthoryear{Pezzulli, Fraternali, Boissier  \& Muñoz-Mateos}{Pezzulli et~al.}{2015}]{pezzulli_instantaneous_2015}
Pezzulli G.,  Fraternali F.,  Boissier S.,   Muñoz-Mateos J.~C.,  2015, \mn@doi [Monthly Notices of the Royal Astronomical Society] {10.1093/mnras/stv1077}, 451, 2324

\bibitem[\protect\citeauthoryear{Popping, Somerville  \& Galametz}{Popping et~al.}{2017}]{popping_dust_2017}
Popping G.,  Somerville R.~S.,   Galametz M.,  2017, \mn@doi [Monthly Notices of the Royal Astronomical Society] {10.1093/mnras/stx1545}, 471, 3152

\bibitem[\protect\citeauthoryear{Popping et~al.,}{Popping et~al.}{2022}]{popping_dust-continuum_2022}
Popping G.,  et~al., 2022, \mn@doi [Monthly Notices of the Royal Astronomical Society] {10.1093/mnras/stab3312}, 510, 3321

\bibitem[\protect\citeauthoryear{Price et~al.,}{Price et~al.}{2014}]{price_direct_2014}
Price S.~H.,  et~al., 2014, \mn@doi [The Astrophysical Journal] {10.1088/0004-637X/788/1/86}, 788, 86

\bibitem[\protect\citeauthoryear{Puglisi et~al.,}{Puglisi et~al.}{2016}]{puglisi_dust_2016}
Puglisi A.,  et~al., 2016, \mn@doi [Astronomy and Astrophysics] {10.1051/0004-6361/201526782}, 586, A83

\bibitem[\protect\citeauthoryear{Rawle et~al.,}{Rawle et~al.}{2022}]{rawle_-flight_2022}
Rawle T.~D.,  et~al., 2022, In-flight performance of the {NIRSpec} {Micro} {Shutter} {Array}, \mn@doi{10.48550/arXiv.2208.04673}, \url {http://arxiv.org/abs/2208.04673}

\bibitem[\protect\citeauthoryear{Reddy, Pettini, Steidel, Shapley, Erb  \& Law}{Reddy et~al.}{2012}]{reddy_characteristic_2012}
Reddy N.~A.,  Pettini M.,  Steidel C.~C.,  Shapley A.~E.,  Erb D.~K.,   Law D.~R.,  2012, \mn@doi [The Astrophysical Journal] {10.1088/0004-637X/754/1/25}, 754, 25

\bibitem[\protect\citeauthoryear{Reddy et~al.,}{Reddy et~al.}{2015}]{reddy_mosdef_2015}
Reddy N.~A.,  et~al., 2015, \mn@doi [The Astrophysical Journal] {10.1088/0004-637X/806/2/259}, 806, 259

\bibitem[\protect\citeauthoryear{Reddy et~al.,}{Reddy et~al.}{2020}]{reddy_mosdef_2020}
Reddy N.~A.,  et~al., 2020, \mn@doi [The Astrophysical Journal] {10.3847/1538-4357/abb674}, 902, 123

\bibitem[\protect\citeauthoryear{Salim \& Narayanan}{Salim \& Narayanan}{2020}]{salim_dust_2020}
Salim S.,  Narayanan D.,  2020, \mn@doi [Annual Review of Astronomy and Astrophysics] {10.1146/annurev-astro-032620-021933}, 58, 529

\bibitem[\protect\citeauthoryear{Salim, Boquien  \& Lee}{Salim et~al.}{2018}]{salim_dust_2018}
Salim S.,  Boquien M.,   Lee J.~C.,  2018, \mn@doi [The Astrophysical Journal] {10.3847/1538-4357/aabf3c}, 859, 11

\bibitem[\protect\citeauthoryear{Salmon et~al.,}{Salmon et~al.}{2016}]{salmon_breaking_2016}
Salmon B.,  et~al., 2016, \mn@doi [The Astrophysical Journal] {10.3847/0004-637X/827/1/20}, 827, 20

\bibitem[\protect\citeauthoryear{Sanders et~al.,}{Sanders et~al.}{2024}]{sanders_aurora_2024}
Sanders R.~L.,  et~al., 2024, The {AURORA} {Survey}: {The} {Nebular} {Attenuation} {Curve} of a {Galaxy} at z=4.41 from {Ultraviolet} to {Near}-{Infrared} {Wavelengths}, \mn@doi{10.48550/arXiv.2408.05273}, \url {http://arxiv.org/abs/2408.05273}

\bibitem[\protect\citeauthoryear{Sandles et~al.,}{Sandles et~al.}{2024}]{sandles_jades_2024}
Sandles L.,  et~al., 2024, \mn@doi [Astronomy and Astrophysics] {10.1051/0004-6361/202347119}, 691, A305

\bibitem[\protect\citeauthoryear{Sarangi, Matsuura  \& Micelotta}{Sarangi et~al.}{2018}]{sarangi_dust_2018}
Sarangi A.,  Matsuura M.,   Micelotta E.~R.,  2018, \mn@doi [Space Science Reviews] {10.1007/s11214-018-0492-7}, 214, 63

\bibitem[\protect\citeauthoryear{Schneider \& Maiolino}{Schneider \& Maiolino}{2024}]{schneider_formation_2024}
Schneider R.,  Maiolino R.,  2024, \mn@doi [Astronomy and Astrophysics Review] {10.1007/s00159-024-00151-2}, 32, 2

\bibitem[\protect\citeauthoryear{Seon \& Draine}{Seon \& Draine}{2016}]{seon_radiative_2016}
Seon K.-I.,  Draine B.~T.,  2016, \mn@doi [The Astrophysical Journal] {10.3847/1538-4357/833/2/201}, 833, 201

\bibitem[\protect\citeauthoryear{Shen, Vogelsberger, Nelson, Tacchella, Hernquist, Springel, Marinacci  \& Torrey}{Shen et~al.}{2022}]{shen_high-redshift_2022}
Shen X.,  Vogelsberger M.,  Nelson D.,  Tacchella S.,  Hernquist L.,  Springel V.,  Marinacci F.,   Torrey P.,  2022, \mn@doi [Monthly Notices of the Royal Astronomical Society] {10.1093/mnras/stab3794}, 510, 5560

\bibitem[\protect\citeauthoryear{Shivaei, Reddy, Steidel  \& Shapley}{Shivaei et~al.}{2015}]{shivaei_investigating_2015}
Shivaei I.,  Reddy N.~A.,  Steidel C.~C.,   Shapley A.~E.,  2015, \mn@doi [The Astrophysical Journal] {10.1088/0004-637X/804/2/149}, 804, 149

\bibitem[\protect\citeauthoryear{Shivaei et~al.,}{Shivaei et~al.}{2020}]{shivaei_mosdef_2020}
Shivaei I.,  et~al., 2020, \mn@doi [The Astrophysical Journal] {10.3847/1538-4357/aba35e}, 899, 117

\bibitem[\protect\citeauthoryear{Skelton et~al.,}{Skelton et~al.}{2014}]{skelton_3d-hst_2014}
Skelton R.~E.,  et~al., 2014, \mn@doi [The Astrophysical Journal Supplement Series] {10.1088/0067-0049/214/2/24}, 214, 24

\bibitem[\protect\citeauthoryear{Sommovigo et~al.,}{Sommovigo et~al.}{2025}]{sommovigo_learning_2025}
Sommovigo L.,  et~al., 2025, Learning the {Universe}: physically-motivated priors for dust attenuation curves, \mn@doi{10.48550/arXiv.2502.13240}, \url {http://arxiv.org/abs/2502.13240}

\bibitem[\protect\citeauthoryear{Stecher \& Donn}{Stecher \& Donn}{1965}]{stecher_graphite_1965}
Stecher T.~P.,  Donn B.,  1965, \mn@doi [The Astrophysical Journal] {10.1086/148461}, 142, 1681

\bibitem[\protect\citeauthoryear{Steglich, Jäger, Rouillé, Huisken, Mutschke  \& Henning}{Steglich et~al.}{2010}]{steglich_electronic_2010}
Steglich M.,  Jäger C.,  Rouillé G.,  Huisken F.,  Mutschke H.,   Henning T.,  2010, \mn@doi [The Astrophysical Journal Letters] {10.1088/2041-8205/712/1/L16}, 712, L16

\bibitem[\protect\citeauthoryear{Stratta, Gallerani  \& Maiolino}{Stratta et~al.}{2011}]{stratta_is_2011}
Stratta G.,  Gallerani S.,   Maiolino R.,  2011, \mn@doi [Astronomy and Astrophysics] {10.1051/0004-6361/201016414}, 532, A45

\bibitem[\protect\citeauthoryear{Suess, Kriek, Price  \& Barro}{Suess et~al.}{2019}]{suess_half-mass_2019}
Suess K.~A.,  Kriek M.,  Price S.~H.,   Barro G.,  2019, \mn@doi [The Astrophysical Journal] {10.3847/1538-4357/ab1bda}, 877, 103

\bibitem[\protect\citeauthoryear{Suess et~al.,}{Suess et~al.}{2022}]{suess_rest-frame_2022}
Suess K.~A.,  et~al., 2022, \mn@doi [The Astrophysical Journal Letters] {10.3847/2041-8213/ac8e06}, 937, L33

\bibitem[\protect\citeauthoryear{Szomoru, Franx, van Dokkum, Trenti, Illingworth, Labbé  \& Oesch}{Szomoru et~al.}{2013}]{szomoru_stellar_2013}
Szomoru D.,  Franx M.,  van Dokkum P.~G.,  Trenti M.,  Illingworth G.~D.,  Labbé I.,   Oesch P.,  2013, \mn@doi [The Astrophysical Journal] {10.1088/0004-637X/763/2/73}, 763, 73

\bibitem[\protect\citeauthoryear{Tacchella, Dekel, Carollo, Ceverino, DeGraf, Lapiner, Mandelker  \& Primack~Joel}{Tacchella et~al.}{2016a}]{tacchella_confinement_2016}
Tacchella S.,  Dekel A.,  Carollo C.~M.,  Ceverino D.,  DeGraf C.,  Lapiner S.,  Mandelker N.,   Primack~Joel R.,  2016a, \mn@doi [Monthly Notices of the Royal Astronomical Society] {10.1093/mnras/stw131}, 457, 2790

\bibitem[\protect\citeauthoryear{Tacchella, Dekel, Carollo, Ceverino, DeGraf, Lapiner, Mandelker  \& Primack}{Tacchella et~al.}{2016b}]{tacchella_evolution_2016}
Tacchella S.,  Dekel A.,  Carollo C.~M.,  Ceverino D.,  DeGraf C.,  Lapiner S.,  Mandelker N.,   Primack J.~R.,  2016b, \mn@doi [Monthly Notices of the Royal Astronomical Society] {10.1093/mnras/stw303}, 458, 242

\bibitem[\protect\citeauthoryear{Tacchella et~al.,}{Tacchella et~al.}{2018}]{tacchella_dust_2018}
Tacchella S.,  et~al., 2018, \mn@doi [The Astrophysical Journal] {10.3847/1538-4357/aabf8b}, 859, 56

\bibitem[\protect\citeauthoryear{Tacchella et~al.,}{Tacchella et~al.}{2022a}]{tacchella_h_2022}
Tacchella S.,  et~al., 2022a, \mn@doi [Monthly Notices of the Royal Astronomical Society] {10.1093/mnras/stac818}, 513, 2904

\bibitem[\protect\citeauthoryear{Tacchella et~al.,}{Tacchella et~al.}{2022b}]{tacchella_fast_2022}
Tacchella S.,  et~al., 2022b, \mn@doi [The Astrophysical Journal] {10.3847/1538-4357/ac449b}, 926, 134

\bibitem[\protect\citeauthoryear{Tacchella et~al.,}{Tacchella et~al.}{2022c}]{tacchella_stellar_2022}
Tacchella S.,  et~al., 2022c, \mn@doi [The Astrophysical Journal] {10.3847/1538-4357/ac4cad}, 927, 170

\bibitem[\protect\citeauthoryear{Trayford, Lagos, Robotham  \& Obreschkow}{Trayford et~al.}{2020}]{trayford_fade_2020}
Trayford J.~W.,  Lagos C. d.~P.,  Robotham A. S.~G.,   Obreschkow D.,  2020, \mn@doi [Monthly Notices of the Royal Astronomical Society] {10.1093/mnras/stz3234}, 491, 3937

\bibitem[\protect\citeauthoryear{Vincent \& Ryden}{Vincent \& Ryden}{2005}]{vincent_dependence_2005}
Vincent R.~A.,  Ryden B.~S.,  2005, \mn@doi [The Astrophysical Journal] {10.1086/428765}, 623, 137

\bibitem[\protect\citeauthoryear{Wang et~al.,}{Wang et~al.}{2021}]{wang_luminous_2021}
Wang F.,  et~al., 2021, \mn@doi [The Astrophysical Journal Letters] {10.3847/2041-8213/abd8c6}, 907, L1

\bibitem[\protect\citeauthoryear{Ward et~al.,}{Ward et~al.}{2024}]{ward_evolution_2024}
Ward E.,  et~al., 2024, \mn@doi [The Astrophysical Journal] {10.3847/1538-4357/ad20ed}, 962, 176

\bibitem[\protect\citeauthoryear{Weingartner \& Draine}{Weingartner \& Draine}{2001}]{weingartner_dust_2001}
Weingartner J.~C.,  Draine B.~T.,  2001, \mn@doi [The Astrophysical Journal] {10.1086/318651}, 548, 296

\bibitem[\protect\citeauthoryear{Wild, Charlot, Brinchmann, Heckman, Vince, Pacifici  \& Chevallard}{Wild et~al.}{2011}]{wild_empirical_2011}
Wild V.,  Charlot S.,  Brinchmann J.,  Heckman T.,  Vince O.,  Pacifici C.,   Chevallard J.,  2011, \mn@doi [Monthly Notices of the Royal Astronomical Society] {10.1111/j.1365-2966.2011.19367.x}, 417, 1760

\bibitem[\protect\citeauthoryear{Witstok et~al.,}{Witstok et~al.}{2023}]{witstok_carbonaceous_2023}
Witstok J.,  et~al., 2023, \mn@doi [Nature] {10.1038/s41586-023-06413-w}, 621, 267

\bibitem[\protect\citeauthoryear{Wuyts et~al.,}{Wuyts et~al.}{2012}]{wuyts_smoother_2012}
Wuyts S.,  et~al., 2012, \mn@doi [The Astrophysical Journal] {10.1088/0004-637X/753/2/114}, 753, 114

\bibitem[\protect\citeauthoryear{Yang, Roberts-Borsani, Treu, Birrer, Morishita  \& Bradač}{Yang et~al.}{2021}]{yang_evolution_2021}
Yang L.,  Roberts-Borsani G.,  Treu T.,  Birrer S.,  Morishita T.,   Bradač M.,  2021, \mn@doi [Monthly Notices of the Royal Astronomical Society] {10.1093/mnras/staa3713}, 501, 1028

\bibitem[\protect\citeauthoryear{Yang et~al.,}{Yang et~al.}{2025}]{yang_cosmos-web_2025}
Yang L.,  et~al., 2025, {COSMOS}-{Web}: {Unraveling} the {Evolution} of {Galaxy} {Size} and {Related} {Properties} at \$2{\textless}z{\textless}10\$, \mn@doi{10.48550/arXiv.2504.07185}, \url {http://arxiv.org/abs/2504.07185}

\bibitem[\protect\citeauthoryear{Zhang et~al.,}{Zhang et~al.}{2019}]{zhang_evolution_2019}
Zhang H.,  et~al., 2019, \mn@doi [Monthly Notices of the Royal Astronomical Society] {10.1093/mnras/stz339}, 484, 5170

\bibitem[\protect\citeauthoryear{Zhang et~al.,}{Zhang et~al.}{2023}]{zhang_dust_2023}
Zhang J.,  et~al., 2023, \mn@doi [Monthly Notices of the Royal Astronomical Society] {10.1093/mnras/stad2066}, 524, 4128

\bibitem[\protect\citeauthoryear{Zuckerman, Belli, Leja  \& Tacchella}{Zuckerman et~al.}{2021}]{zuckerman_reproducing_2021}
Zuckerman L.~D.,  Belli S.,  Leja J.,   Tacchella S.,  2021, \mn@doi [The Astrophysical Journal] {10.3847/2041-8213/ac3831}, 922, L32

\bibitem[\protect\citeauthoryear{da Cunha, Eminian, Charlot  \& Blaizot}{da~Cunha et~al.}{2010}]{da_cunha_new_2010}
da Cunha E.,  Eminian C.,  Charlot S.,   Blaizot J.,  2010, \mn@doi [Monthly Notices of the Royal Astronomical Society] {10.1111/j.1365-2966.2010.16344.x}, 403, 1894

\bibitem[\protect\citeauthoryear{van~der Wel et~al.,}{van~der Wel et~al.}{2014}]{van_der_wel_geometry_2014}
van~der Wel A.,  et~al., 2014, \mn@doi [The Astrophysical Journal] {10.1088/2041-8205/792/1/L6}, 792, L6

\makeatother
\end{thebibliography}




\appendix

\section{FSPS Toy Model}

\subsection{Model Attenuation Law Variation} \label{App:fiducial}

To model the SFH and explore how it affects the dust attenuation law, following the two-component dust model, we only consider the number of stars older or younger than $10$ Myr. Hence, for simplicity, we used a flat step SFH with constant SFR from birth (1/10th the age of the universe after the big bang) to $10$ Myr ago, then a different SFR from $10$ Myr ago to the present. The total mass of this galaxy was fixed to 10$^{10}$ M$_{\odot}$. Using this parameterisation, the ratio of stellar mass formed in the last $10$ Myr over the total stellar mass, f$_{10}$, can vary the SFH. 

To determine the attenuation law for each of these galaxy instances, we synthesised the spectra through FSPS with dust turned on, then turned off, to get the attenuated flux ($F_{\lambda, \rm dust}$) and intrinsic flux ($F_{\lambda, 0}$) respectively, and get the spectral attenuation ($A_{\lambda}$) from $A_{\lambda} = 2.5\log(F_{\lambda, 0}/F_{\lambda, \rm dust})$.

\subsection{Young Stellar Population Flux Contribution}\label{App:fl10}

\begin{figure}
\centerline{\includegraphics[width=.5\textwidth]{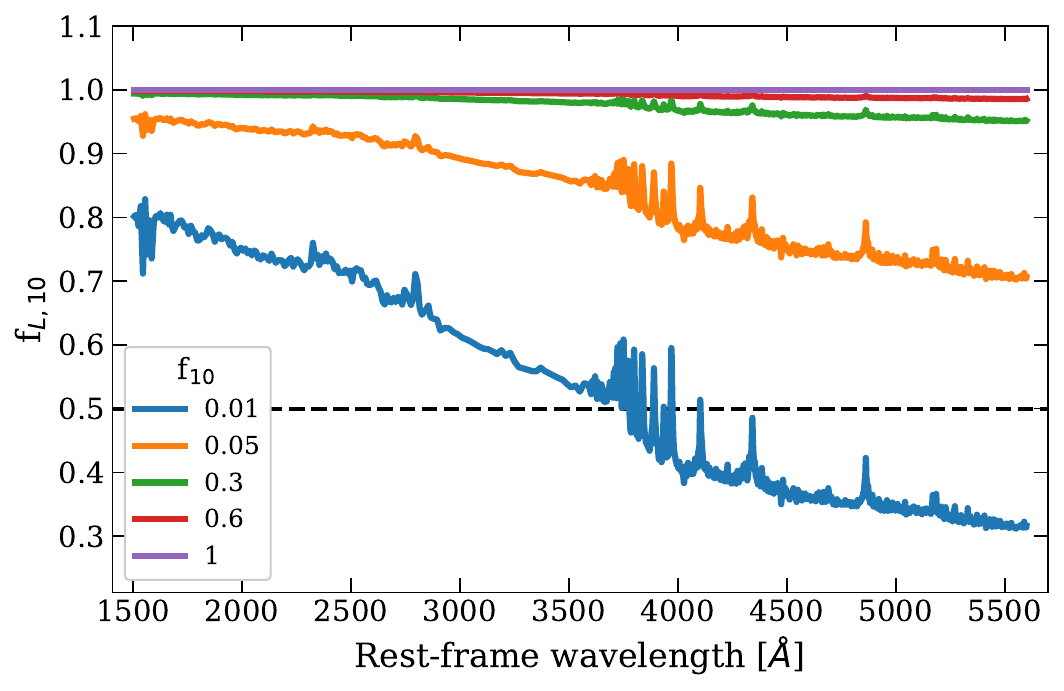}}
\medskip
\caption{Ratio of flux from the young stars ($<10$ Myr) to the total flux from a model galaxy in FSPS. The young stars dominate the UV regime even when the young stars make up $1\%$ of the total stellar mass ($f_{10}=0.01$. The spikes in the flux ratio at wavelength $\lambda>3500$\AA{} at low values of $f_{10}$ are due to stellar absorption lines from older stars, which contribute significantly to the spectrum at optical wavelengths. Over short wavelength ranges, the relative flux contribution between the stellar populations will vary.} \label{f:fsps_f_L_10}
\end{figure}

Since the way the dust from the birth clouds and the ISM affect the overall attenuation law is non-trivial, we show in Figure~\ref{f:fsps_f_L_10} the spectral flux ratio from the young ($<10$ Myr) stars to the total population ($f_{L,10}$), as the mass ratio of these two populations ($f_{10}$) varies. To calculate how much flux the young ($<10$ Myr) and old ($>10$ Myr) stellar populations contribute to the overall SED at different wavelengths, we model the spectra of a galaxy in FSPS. We set the diffused ISM dust and the birth cloud dust to have zero optical depth, i.e., there is no dust, so that we can study the intrinsic spectra of the stars. The mass ratio of young stars to the total stellar population, $f_{10}$, varies from $0.1-1$. A flat step SFH is used, as is described in Appendix~\ref{App:fiducial}, to determine the spectrum of the full galaxy. We then set the SFR for $t>10$Mr to be zero to measure the spectrum for just the young stars at each value of $f_{10}$. The ratio of these spectra is $f_{L,10}$.

Since the young stars emit mainly in the UV, and the older stars mainly at the redder wavelengths, we expect to see $f_{L,10}$ increase as the wavelengths become bluer. As $f_{10}$ increases, the relative flux from the young population increases; however, the young population dominates the UV region even at low values of $f_{10}$. The Balmer jump can be seen around 3645 \AA\, where the older stars start emitting more strongly redward of this jump, reducing the flux ratio from the young stars. Also, past the Balmer jump, spikes in the flux ratio can be seen, which are due to effects such as the absorption of certain wavelengths of stellar flux in the atmospheres of older stars, causing a sharp change in the relative flux between the populations over a short wavelength range. Here, nebular emission is turned off.

From Figure~\ref{f:fsps_f_L_10}, it can be seen that as the stellar populations change, the flux contribution from each population changes. This, combined with the attenuation laws of the birth cloud dust and the diffuse ISM dust, leads to the total, effective attenuation law. If a galaxy has mostly old stars, only variations in the ISM dust attenuation law will affect the effective attenuation law, since there would be little birth cloud dust. 

\bsp	
\label{lastpage}
\end{document}